\documentclass{IEEEtran}
\usepackage{amsfonts}
\usepackage{mathrsfs}
\usepackage{amssymb,amsmath}
\usepackage[noblocks]{authblk}
\usepackage[compress]{cite}
\usepackage{bm}
\usepackage{algorithm}
\usepackage{algorithmic}
\usepackage{amsmath,amssymb,epsfig,graphics,subfigure}
\usepackage{theorem}
\usepackage{array,color}
\usepackage[compress]{cite}
\usepackage{multirow}
\usepackage{color}

\usepackage{graphics}
\usepackage{epsfig}
\usepackage{graphicx}

\usepackage{graphicx}
\usepackage{amsmath}
\usepackage{array}
\newcommand{\PreserveBackslash}[1]{\let\temp=\\#1\let\\=\temp}
\newcolumntype{C}[1]{>{\PreserveBackslash\centering}p{#1}}
\newcolumntype{R}[1]{>{\PreserveBackslash\raggedleft}p{#1}}
\newcolumntype{L}[1]{>{\PreserveBackslash\raggedright}p{#1}}
\usepackage{tabularx}
\usepackage{multicol}
\usepackage{booktabs}
\usepackage[Symbol]{upgreek}
\usepackage{subfigure}
\usepackage{bm}
\usepackage{cite}
\usepackage{stfloats}
\usepackage{balance}

\newtheorem{lemma}{Lemma}

\theoremheaderfont{\normalfont\bfseries}

\begin{document}

\title{\huge Polarization Sensitive Array Based Physical-Layer Security}

\author{Shiqi Gong, Chengwen Xing, Sheng Chen,~\emph{Fellow, IEEE}, and  Zesong Fei  %
\thanks{S. Gong, C. Xing and Z. Fei are with School of Information and Electronics,
 Beijing Institute of Technology, Beijing 100081, China (E-mails: gsqyx@163.com,
 xingchengwen@gmail.com, feizesong@bit.edu.cn).}%
\thanks{S. Chen is with Electronics and Computer Science, University of Southampton,
 Southampton SO17 1BJ, U.K. (E-mail: sqc@ecs.soton.ac.uk), and also with King Abdulaziz
 University, Jeddah 21589, Saudi Arabia}%
\vspace*{-5mm}
}

\maketitle

\begin{abstract}
 We propose a framework exploiting the polarization sensitive array ({PSA}) to improve
 the physical layer security of wireless communications. Specifically, the
 polarization difference among signals is utilized to improve the secrecy rate of
 wireless communications, especially when these signals are spatially indistinguishable.
 We firstly investigate the {PSA} based secure communications for point-to-point wireless
 systems from the perspectives of both total power minimization and secrecy rate
 maximization. We then apply the {PSA} based secure beamforming designs to relaying
 networks. The secrecy rate maximization for relaying networks is discussed in detail
 under both the perfect channel state information and the polarization sensitive array
 pointing error. In the later case, a robust scheme to achieve secure communications
 for relaying networks is proposed. Simulation results show that the proposed {PSA}  based
 algorithms achieve lower total power consumption and better security performance
 compared to the conventional scalar array designs, especially under challenging
 environments where all received signals at destination are difficult to distinguish in
 the spatial domain.
\end{abstract}


\section{Introduction}\label{S1}

 The issue of information security in wireless networks has attracted extensive attention
 in recent years considering the openness of wireless links \cite{1,2}. Traditionally,
 encryption techniques are utilized to ensure secure communications, which are generally
 applied in the upper layer of network and have a high design complexity \cite{3}.
 Therefore, an intrinsic approach exploring the characteristics of wireless fading
 channels to improve information security emerges as a prominent technique, which is
 referred to as the physical layer security \cite{4}. The fundamental theory for physical
 layer security was firstly established by Shannon \cite{shannon5}. Following Shannon's
 work, Wyner \cite{wyner6} introduced the famous wiretap channel model and further defined
 the channel secrecy capacity. The work \cite{7} proposed a Gaussian degraded wiretap
 channel which is widely used to model the wireless propagation environment.

 Based on these pioneering theoretical concepts, a large amount of literature focusing
 on various design aspects of secure communications have sprung up. By applying multiple
 antennas at communication nodes to exploit spatial freedom, these researches aimed to
 significantly improve the physical layer security of wireless networks
 \cite{AN9,CJ10,MIMO11,AN11}. For example, an artificial noise scheme was proposed
 for wiretap channels in \cite{AN9} to study the impact of antenna selection on security
 performance of multi-input multi-output (MIMO) two-way relaying networks. The work
 \cite{CJ10} introduced an effective method called cooperative jamming to confuse the
 eavesdropper deliberately. With the aid of the game theory, a collaborative
 physical-layer security transmission scheme was designed in \cite{MIMO11} to effectively
 balance the security performance among different links. All these works however assume
 that the wireless channels are ideally Rayleigh distributed, which ignores the influence
 of array directivity and correlation.  A technique known as the directional modulation
 was also investigated to realize  secure communications. In the work \cite{DM1}, the
 directional modulation technique was applied to the phased  array to offer security.
 Specifically, by shifting each array element's phase appropriately, the desired symbol
 phase and amplitude in a given direction is generated. The study \cite{DM2} on the
 other hand adopted the directional modulation technique to enhance the security of
 multi-user MIMO systems. Different from the standard secrecy rate optimization, the secure
 communications of multi-user MIMO systems are achieved by increasing the symbol error rate
 at the eavesdropper. It can be seen that the directional modulation technique designs the
 weighting coefficients of the phased array. As will be shown, our polarization sensitive
 array (PSA) based technique designs the spatial pointing of each antenna to effectively
 extract the signals' polarization information for realizing secure communications.

 Generally, the polarization status, similar to the amplitude and phase, is a feature
 of the signal. Many researches have indicated that the direction-finding performance
 and short-wave communication quality can be improved by means of the polarization
 difference among signals \cite{plo14}. However, in many practical communication
 scenarios, such as radar and electronic reconnaissance, the conventional scalar array
 (CSA) is widely deployed. In essence, the CSA is the uniformly spaced linear array with
 the same spatial properties in all its array elements. Generally, CSA is blind to the
 polarization status of signal and sensitive to the array aperture and signal wavelength
 \cite{scalar15}. Worse still, in some specific array alignment, a CSA may present the
 morbid response to the polarization status of signal. Different from the CSA, the {PSA}
 consists of a certain number of antennas with different spatial pointings, which can be
 utilized to extract the signal information more meticulously and comprehensively in a
 vector way \cite{plo16, pp2}. The spatial pointings of the PSA offer extra design
 degrees of freedom for physical layer security of wireless networks. In most practical
 wireless networks, jammer is typically introduced to effectively interfere with the eavesdropper,
 but it simultaneously causes the interference to the destination. When the jammer signal
 has approximately the same spatial properties as the source, the CSA based destination
 beamforming optimization is unable to suppress the interference, as it can only rely on
 the signals' spatial characteristics. By contrast, since different polarization information
 can  be extracted by the spatial pointings of PSA, the PSA based destination beamforming
 optimization is capable of suppressing the interference effectively, even when the signals
 are indistinguishable in the spatial domain. Therefore, utilizing the PSA to realize
 secure communications for wireless networks can achieve superior performance over the
 CSA design.

 However, most existing  PSA related works focus on the problem of estimating the signal's
 direction of arrival (DOA). In \cite{plo19}, a two-step maximum-likelihood signal
 estimation procedure was developed under the PSA. Based on the sparse polarization sensor
 measurements, the DOA estimation of the transmitted signal was conducted in \cite{plo199}.
 There also exist some works specifically related to the optimization of dual-polarization
 array to enhance the system capacity. Compared to the single polarization array, the
 orthogonal dual polarization antenna can enhance MIMO spatial multiplexing gain remarkably
 by means of the eigenvalue ratio decomposition \cite{dual21}. The study \cite{dual22}
 designed a linear-polarized dual-polarization frequency reuse system to increase spectrum
 utilization and further improve the system capacity, while the work \cite{dual23} compared
 three different transmission schemes for MIMO networks to achieve the maximum diversity
 under a dual-polarization channel model. All these works do not consider utilizing {PSA} to
 enhance secure communications.

 Against the above background, this paper investigates the {PSA} based secure transmission
 strategy for wireless networks. Specifically, we first consider the { PSA} based secure
 communications for the point-to-point single-input multi-output (SIMO) network with the
 aid of jammer. In this case, the secure beamforming is firstly designed aiming at
 minimizing the total transmit power subject to the secrecy rate requirements. Then the
 secrecy rate maximization scheme is proposed to improve the secrecy capacity of SIMO
 network as much as possible. Further extending our research into the more complicated
 scenario where the relay is employed to enlarge the communication coverage of source nodes,
 we consider the secrecy rate maximization under both perfect channel state information
 (CSI) and imperfect  PSA pointing, respectively. It is worth noting that convex optimization
 techniques \cite{cvxx} can be utilized to solve the optimization problems formulated in this
 paper effectively.

 The rest of the paper is organized as follows. In Section~\ref{S2}, the system model of
 {PSA} is briefly introduced. In Section~\ref{S3}, the point-to-point secrecy beamforming
 is designed for SIMO networks, while the one-way relaying network is considered in
 Section~\ref{S4}, where the corresponding secrecy rate optimization problems are
 formulated. Section~\ref{S5} presents the simulation results, and our conclusions are
 given in Section~\ref{S6}.

 The normal-faced lower-case letters denote scalars, while bold-faced lower-case and upper-case
 letters stand for vectors and matrices, respectively. $| ~ |$ denotes the absolute value
 and $\| ~ \|$ denotes the Euclidean norm, while $( ~ )^{\rm \ast}$, $( ~ )^{\rm T}$,
 $( ~ )^{\rm H}$ and $( ~ )^{-1}$ represent the conjugate, transpose, conjugate transpose
 and inverse operators, respectively. An optimal solution is marked by $^{\star}$, while
 $\text{tr}(~)$ and $\text{rank}(~)$ denote the trace and rank of matrix, respectively. The
 $n$th row of matrix $\bm{A}$ is given by $\bm{A}[n,:]$, and the $n$th-row and $m$th-column
 element of $\bm{A}$ is $\bm{A}[n,m]$. $\bm{A}\succeq 0$ means that $\bm{A}$ is a positive
 semidefinite matrix. The vector stacking operator $\text{vec}( ~ )$ stacks the columns of
 a matrix on top of one another, and $\text{diag}\big\{u_1,\cdots ,u_N\big\}$ is the diagonal
 matrix with the diagonal elements $u_1,\cdots ,u_N$. $\bm{I}_N$ is the $N\times N$ identity
 matrix, and $\mathbf{0}_{n\times m}$ is the $n\times m$ matrix with all zero elements.
 $\bm{a}\sim{\cal{CN}}(\bm{0},\sigma^2\bm{I})$ means that $\bm{a}$ is a complex Gaussian
 distributed random vector with the zero mean vector $\bm{0}$ and the covariance matrix
 $\sigma^2\bm{I}$, while $\textsf{E}\{ ~\}$ is the expectation operator. The determinant
 operation is denoted by $\det ( ~ )$, and $\otimes$ denotes the Kronecker product. Finally,
 $\textsf{j}=\sqrt{-1}$, and $[a]^{+}=\max\{0,a\}$.

\begin{figure}[hp!]
\vspace*{-4mm}
\begin{center}
 \includegraphics[width=0.48\columnwidth,height=0.39\columnwidth]{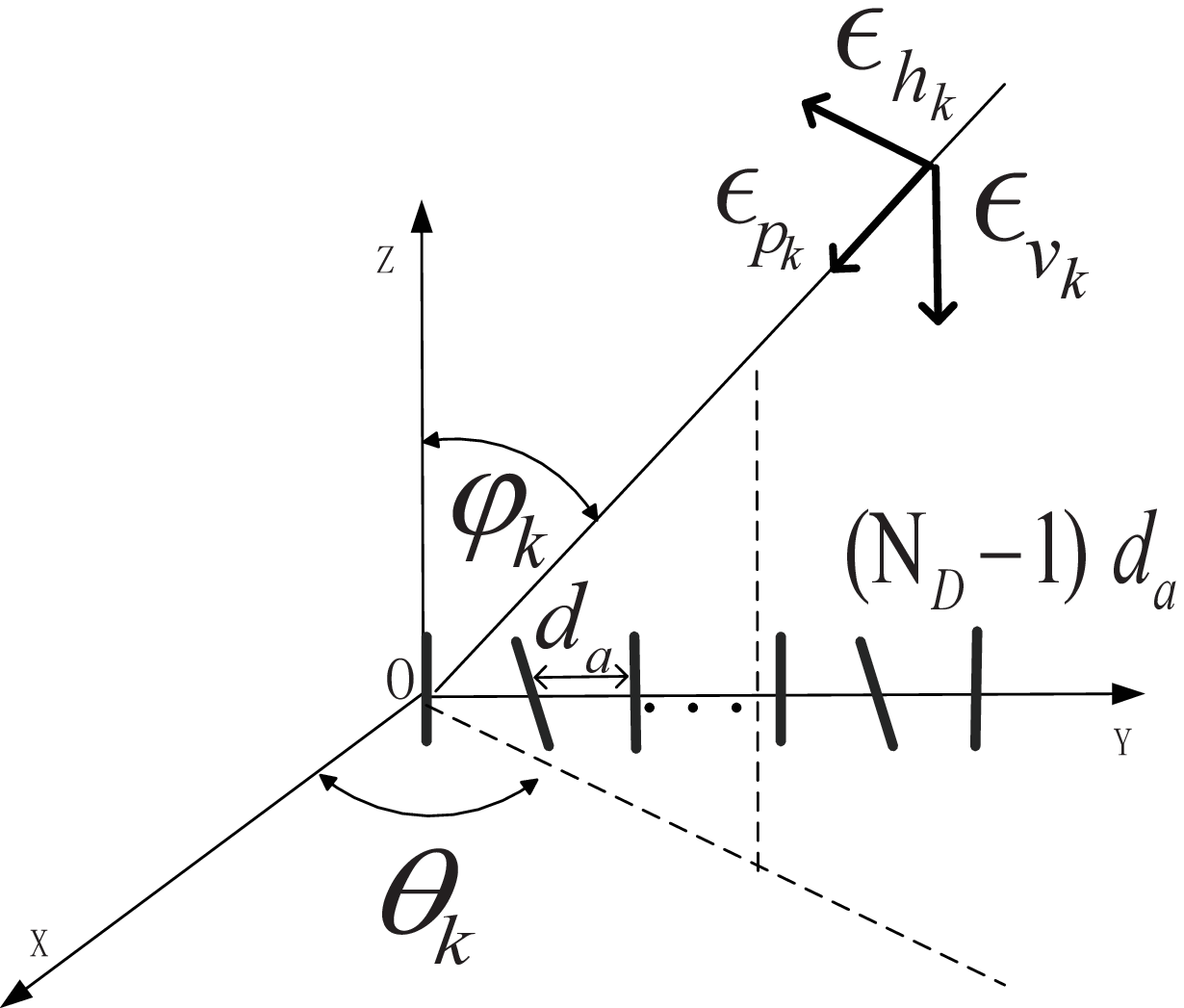}
 \includegraphics[width=0.48\columnwidth,height=0.39\columnwidth]{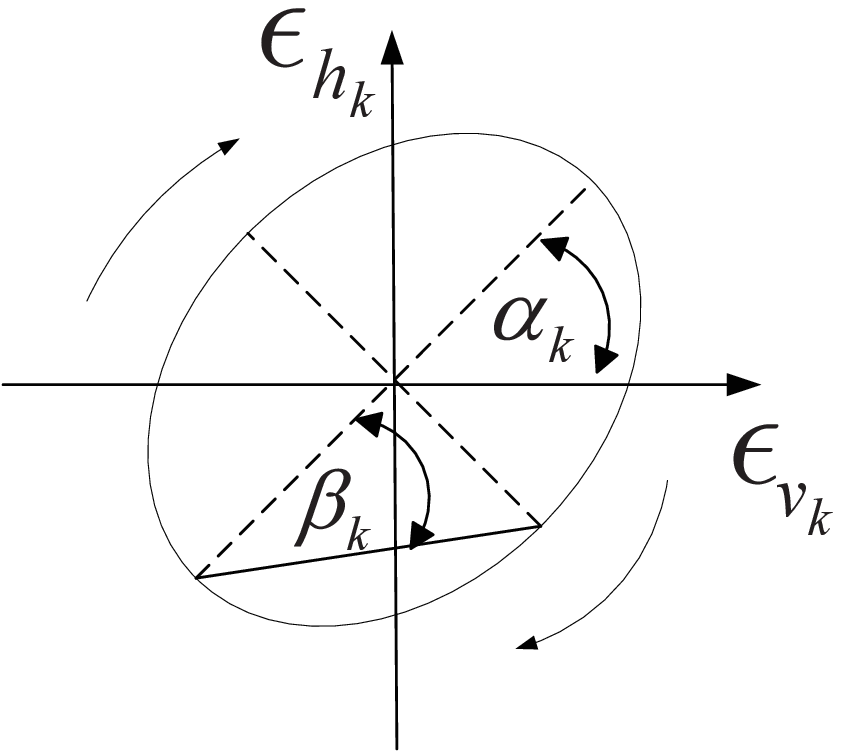}
\end{center}
\vspace*{-5mm}
\begin{center}{\small (a)}\hspace*{35mm}{\small (b)}\end{center}
\vspace*{-5mm}
\caption{Polarization sensitive array model: (a)~the uniform linear crossed dipole
 array with $N_D$ antennas, and (b)~the polarized ellipse of EM signal.}
\label{FIG1}
\vspace*{-2mm}
\end{figure}

\section{Polarization Sensitive Array System Model}\label{S2}

 Without loss of generality,  we assume that a total of $N_D$ antennas are located
 in the $y$-axis and the distance $d_a$ between the adjacent antennas is half
 wavelength, as illustrated in Fig.\,\ref{FIG1}\,(a). Here two plane electromagnetic
 (EM) signals are considered, i.e., the desired EM signal $s_d$ and the jamming EM
 signal $s_j$. They arrive at the $N_D$ antennas of the {PSA} from different incident
 angles. As is well known, the EM wave is traveling in a single direction, where the
 electric component and the magnetic component are perpendicular to each other as well
 as perpendicular to this propagation direction. Taking the electric component as an
 example, we  define the transverse electric field vectors of the EM signal $s_k$ as
\begin{align}\label{eq1}
 {\bm{e}}_{s_k}(t) =& e_{h_k}(t)\bm{\epsilon}_{h_k} + e_{v_k}(t)\bm{\epsilon}_{v_k} ,\, k=d, j ,
\end{align}
 where $e_{h_k}(t)$ and $e_{v_k}(t)$ are the electric field projections on the
 $\bm{\epsilon}_{h_k}$ and $\bm{\epsilon}_{v_k}$ directions, respectively. As a
 result, the magnetic field biases of the EM signal are $\bm{\epsilon}_{v_k}$ and
 $-\bm{\epsilon}_{h_k}$, respectively, for keeping the orthogonality \cite{ele2}.
 Furthermore, it is assumed that the EM signals are completely polarized signals which
 means that the time varying $e_{h_k}(t)$ and $e_{v_k}(t)$ can be formulated as an
 ellipse. As described in Fig.\,\ref{FIG1}\,(b), $\alpha_k$ and $\beta_k$ are the
 polarization orientation and ellipse angle, respectively, which represent the
 track of the EM signal's electric vector and are thereafter called the POA for short.
 According to the EM theory \cite{ele2,plo199,dual21,dual22,dual23,EE,ele3}, we can
 express the EM signal in a vector form with its DOA $(\theta_k,\varphi_k)$ and POA
 $(\alpha_k,\beta_k)$ as follows
\begin{align}\label{eq2}
 \widehat{\bm{s}}_k \!\!=\!\!\bm{\Xi}\big(\theta_k,\varphi_k\big) \bm{R}\big(\alpha_k\big)
 \bm{\ell}\big(\beta_k\big)
 \!\!=\!\!\big[\widehat{s}_k(1) \cdots \widehat{s}_k(6)
 \big]^{\rm T} ,\, k\!\!=\!\!d, j ,
\end{align}
 where $\theta_k$ and $\varphi_k$ are the azimuth and elevation angles of the EM signal
 ${s}_k$, respectively, while
\begin{align}\label{eq3}
 & \bm{\Xi}\big(\theta_k,\varphi_k\big) \!=\! \left[\!\! \begin{array}{lr} \bm{\epsilon}_{h_k}
  \! & \! \bm{\epsilon}_{v_k} \\ \bm{\epsilon}_{v_k} \! & \! -\bm{\epsilon}_{h_k} \end{array}\!\! \right] \!\!
  = \!\!\left[ \!\! \begin{array}{cc} -\sin\theta_k \! & \! \cos\varphi_k \cos\theta_k \\
  \cos\theta_k \! & \! \cos\varphi_k \sin\theta_k \\ 0 \! & \! -\sin\varphi_k \\
  \cos\varphi_k \cos\theta_k \! & \! \sin \theta_k  \\ \cos\varphi_k \sin\theta_k \! & \! -\cos\theta_k \\
  -\sin\varphi_k \! & \! 0 \end{array} \!\! \right] \!\! ,
\end{align}
\begin{align}\label{eq4}
 & \bm{R}(\alpha_k) \! = \! \left[ \!\! \begin{array}{lr}
  \cos\alpha_k \! & \! -\sin\alpha_k \\ \sin\alpha_k \! & \!\cos\alpha_k \end{array} \!\! \right] \! \text{ and }
 \bm{\ell}(\beta_k) \!\! = \!\!\left[ \! \begin{array}{l}
 \cos\beta_k \\ \textsf{j}\sin\beta_k \end{array}\! \right] \! .
\end{align}
 $\bm{\Xi}\big(\theta_k,\varphi_k\big)$ is the steering matrix of $s_k$, which is composed
 of the electric and magnetic field bases of the EM signal, while $\bm{R}\big(\alpha_k\big)$
 and $\bm{\ell}\big(\beta_k\big)$ are the corresponding rotation and ellipticity matrix of
 $s_k$, respectively, \cite{ele3}.

 In addition to the polarization of EM signals, the antenna polarization should also be
 considered. It is noted that only short dipole antennas are adopted in our work, and thus
 the array magnetic response can be neglected. Besides, the polarization sensitive matrix
 $\bm{P}$ which represents the polarization characteristics of the array is defined by the
 spatial pointing angles of the $N_D$ antennas of the { PSA}, i.e., $\big(\theta^{(e,n)},
 \varphi^{(e,n)}\big)$ for $0\le n\le N_D-1$, where $\theta^{(e,n)}$ and $\varphi^{(e,n)}$
 are the azimuth and elevation pointing angle of the $n$th antenna of the { PSA}, respectively.
 Mathematically, we have
\begin{align}\label{eq5}
 \bm{P}\!\!=\!\!& \left[\bm{P}_e ~ \bm{0}\right] \!\!=\!\! \left[\!\! \begin{array}{cccccc}
 p_{e,x}^{(0)} \! & \! p_{e,y}^{(0)} \! & \! p_{e,z}^{(0)} \! & \! 0 & 0 & 0 \\
 p_{e,x}^{(1)} \! & \! p_{e,y}^{(1)} \! & \! p_{e,z}^{(1)} \! & \! 0 & 0 & 0 \\
 \vdots \! & \! \vdots \! & \! \vdots \! & \! \vdots & \vdots & \vdots \\
 p_{e,x}^{(N_D-1)} \! & \! p_{e,y}^{(N_D-1)} \! & \! p_{e,z}^{(N_D-1)} \! & \! 0 & 0 & 0 \\
 \end{array}\!\! \right] \!\! , \!
\end{align}
 with
\begin{align}\label{eq6}
 \left[\! \begin{array}{l} p_{e,x}^{(n)} \\ p_{e,y}^{(n)} \\ p_{e,z}^{(n)}\end{array}\! \right]
 \! = & G_e \! \left[ \!\! \begin{array}{c}
 \sin\varphi^{(e,n)} \cos\theta^{(e,n)} \\
 \sin\varphi^{(e,n)} \sin\theta^{(e,n)} \\
 \cos\varphi^{(e,n)} \end{array} \!\! \right] \!\! , \, 0\le n\le N_D-1,
\end{align}
 where $G_e$ (generally taking the value of 1) is the antenna gain when the polarization
 status of the EM signal perfectly matches the antenna. Note that the matrix $\bm{0}$
 included in $\bm{P}$ indicates that the array magnetic response is ignored. In addition,
 for the matrix $\bm{P}_e$, we have
\begin{align}\label{eq7}
 \big\|\bm{P}_e[n+1,:]\big\|^2\!=\!1 , \, 0\le n\le N_D-1 .
\end{align}
 It is worth emphasizing that different from \cite{plo199}, where the { PSA} consists of
 the aligned short dipole antennas, each antenna of the {PSA} in our paper is deployed with
 a different spatial pointing angle, which becomes an optimization variable for secure
 communications.

 Furthermore, the space phase matrix $\bm{U}_k$ of the EM signal ${s}_k$ impinging on the { PSA}
 is given by
\begin{align}
 \bm{U}_k =& \text{diag}\big\{u_{k,0},u_{k,1},\cdots ,u_{k,N_D-1}\big\} ,\label{eq8}  \\
 u_{k,n} =& e^{-\textsf{j} 2\pi\big({\bm{\xi}}(\theta_k,\varphi_k){\bm{r}}_n\big)/\lambda_k} \nonumber \\
 =& e^{\textsf{j} \pi n \sin\varphi_k\sin\theta_k}, ~
  k=d,j , ~ 0\le n\le N_D -1 , \label{eq9}
\end{align}
 where ${\bm{\xi}}(\theta_k,\varphi_k)=-\big[\sin\varphi_k\cos\theta_k ~
 \sin\varphi_k\sin\theta_k ~ \cos\varphi_k\big]$ denotes the propagation vector of the
 EM signal $s_k$, ${\bm{r}}_n=[0, nd_a,0]^T$ is the position vector of the $n$th
 polarization antenna, and $\lambda_k$ is the wavelength of $s_k$. Based on \eqref{eq2},
 \eqref{eq5} and \eqref{eq8}, the spatio-polarized manifold for the EM signal
 $\widehat{\bm{s}}_k$ is defined as
\begin{align}\label{eq10}
 \bm{a}_{\theta_k,\varphi_k,\alpha_k,\beta_k} \!\!&= \!\!\bm{U}_k\bm{P}\widehat{\bm{s}}_k \!\!= \!\!
 \bm{U}_k\bm{P}\bm{\Xi}(\theta_k,\varphi_k)\bm{R}(\alpha_k)\bm{\ell}(\beta_k) ,
\end{align}
 for $k=d,j$. For notational convenience, we will simplify
 $\bm{a}_{\theta_k,\varphi_k,\alpha_k,\beta_k}$ as $\bm{a}_k$ in the sequel.

 For the sake of maximizing secrecy rate, the { PSA}'s spatial pointings need to be optimized.
 In order to perform this optimization conveniently, the formulation (\ref{eq10}) is
 rewritten as
\begin{align}\label{eq11}
 \bm{a}_k\!\!=\!\!\bm{U}_k \big[\bm{P}_e ~ \bm{0}\big] \widehat{\bm{s}}_k
   \!\!= \!\!\bm{U}_k \bm{P}_e
 \big[\widehat{s}_k(1) ~ \widehat{s}_k(2) ~ \widehat{s}_k(3)\big]^{\rm T} \!\!=\!\!
 \bm{Q}_k\bm{p},
\end{align}
 for $k=d,j$, where
\begin{align}
 \bm{p} =& \big[ p_{e,x}^{(0)} \cdots p_{e,x}^{(N_D-1)} ~ p_{e,y}^{(0)} \cdots
  p_{e,y}^{(N_D-1)} ~ p_{e,z}^{(0)} \cdots p_{e,z}^{(N_D-1)} \big]^{\rm T} \nonumber \\
 =& \text{vec}\big(\bm{P}_e\big) \in \mathbb{R}^{3N_D} , \label{eq12} \\
 \bm{Q}_k \! =& \! \left[\bm{U}_k\! \otimes\! \widehat{s}_k(1) ~ \bm{U}_k\! \otimes\! \widehat{s}_k(2) ~
 \bm{U}_k\! \otimes\! \widehat{s}_k(3)\right]\!\in\! \mathbb{C}^{N_D\times 3N_D} \! . \! \label{eq13}
\end{align}
 Clearly, the new vector $\bm{p}$ denotes the {PSA}'s spatial pointings and thus
 becomes our optimization variables.

\begin{figure}[hp!]
\vspace*{-2mm}
\begin{center}
\includegraphics[width=0.49\columnwidth,height=0.42\columnwidth]{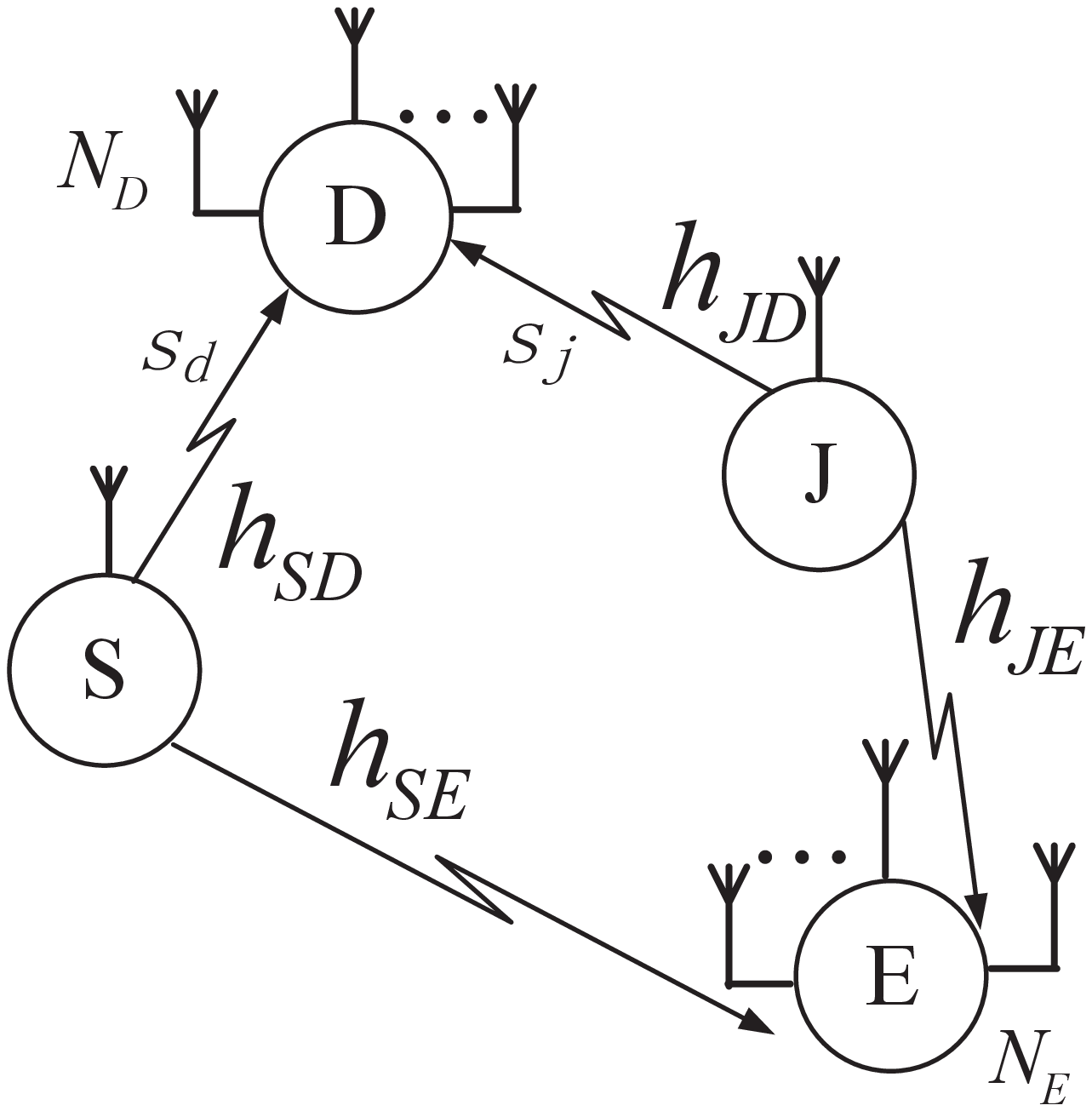}
\includegraphics[width=0.49\columnwidth,height=0.42\columnwidth]{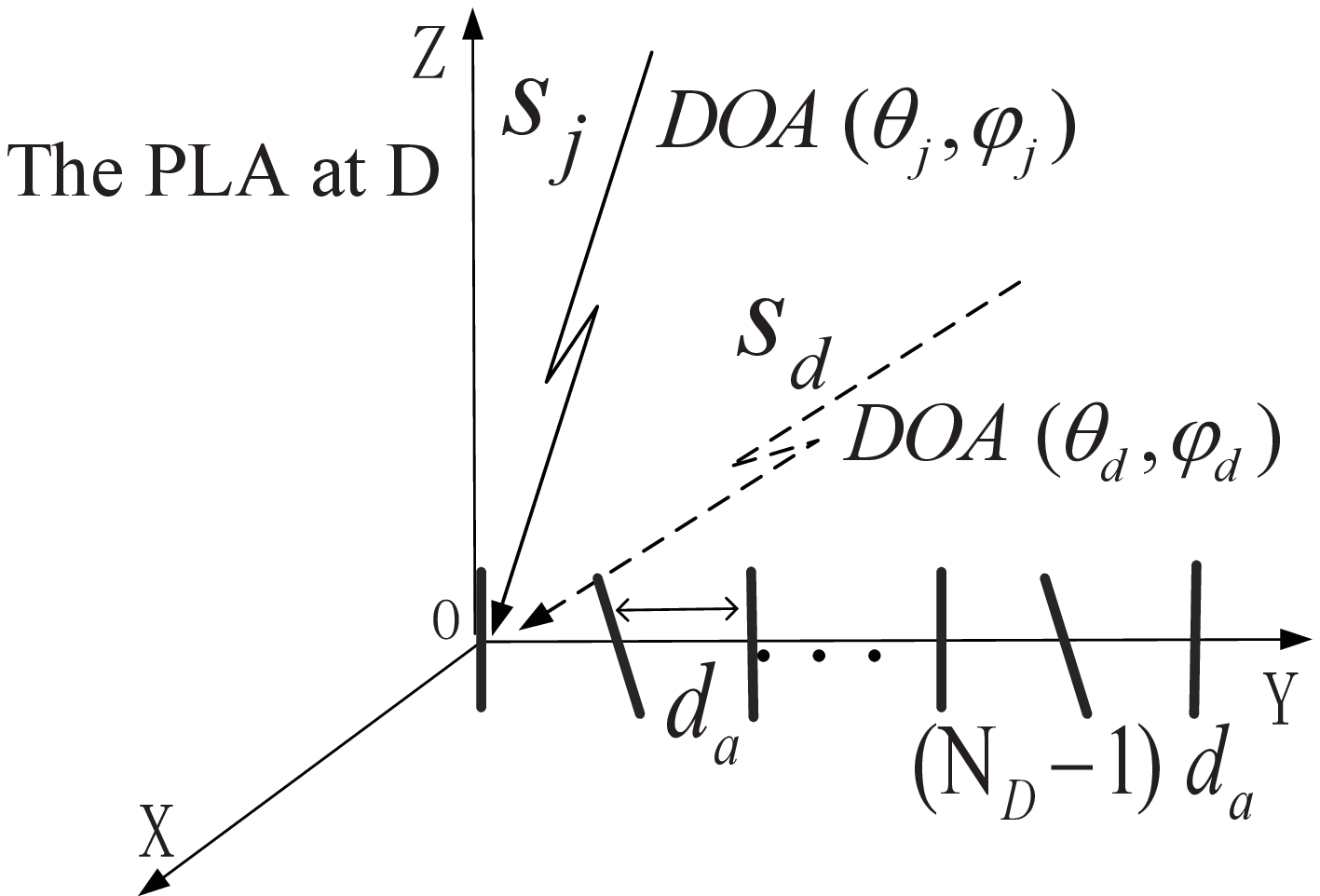}
\end{center}
\vspace*{-5mm}
\caption{A point-to-point SIMO network with the polarization sensitive array having $N_D$
 antennas at destination.}
\label{FIG2}
\vspace*{-4mm}
\end{figure}

\section{Point-to-Point Secrecy Beamforming Design}\label{S3}

 We consider the simplest but most representative wiretap channel as a source, a
 destination and an eavesdropper. In most cases, the capacity of the wiretap channel is
 higher than the main channel owing to the concealment and intention of the eavesdropper.
 In order to realize  secure communications, we introduce a jammer to disturb the
 eavesdropper sufficiently. {In this four-terminal network as depicted in Fig.\,\ref{FIG2}, the source $S$ and jammer $J$ are equipped with single antenna, while the eavesdropper
 $E$ and the  destination $D$  are equipped with $N_E$ and $N_D$ antennas, respectively.  More importantly,  in our work,  the  $N_D$-antenna {PSA} at destination $D$ is assumed instead of the conventional CSA to fully show the advantage of PSA for secure communications }{\footnote{{Our work can easily be extended to
 the more general case, where the eavesdropper is also equipped with the PSA of $N_E$ antennas.
 In fact, in this case, all our designs and algorithms remains applicable and effective.
 Due to the space limitation, the detailed discussions are omitted here.}}}.

\begin{figure*}[tp]\setcounter{equation}{17}
\vspace*{-1mm}
\begin{align}\label{eqr1} 
 & \Big(\big(P_J\bm{h}_{JE}\bm{h}_{JE}^{\rm H}+\sigma_e^2\bm{I}_{N_E}\big)^{-1}\bm{h}_{SE}
  \bm{h}_{SE}^{\rm H}\Big)\Big(\big(P_J\bm{h}_{JE}\bm{h}_{JE}^{\rm H}+\sigma_e^2\bm{I}_{N_E}\big)^{-1}
  \bm{h}_{SE}\Big) = \nonumber \\
 & \hspace*{20mm}\underbrace{\Big(\bm{h}_{SE}^{\rm H}\big(P_J\bm{h}_{JE}\bm{h}_{JE}^{\rm H}+\sigma_e^2
  \bm{I}_{N_E}\big)^{-1}\bm{h}_{SE}\Big)}_{
  \lambda_{\max}}\underbrace{\Big(\big(P_J\bm{h}_{JE}\bm{h}_{JE}^{\rm H}
  +\sigma_e^2\bm{I}_{N_E}\big)^{-1}\bm{h}_{SE}\Big)}_{\bm{\vartheta}_{\max}} .
\end{align}
\hrulefill
\vspace*{-4mm}
\end{figure*}

 Let $\bm{h}_{mn}\sim \mathcal{CN}(\bm{0},\sigma_h^2\bm{I})$ be the channel gain vector
 from node $m$ to node $n$, where $m=S,J$ and $n=E$. Furthermore, we assume far field
 communications related to destination $D$, and we denote $h_{SD}$ and $h_{JD}$ as the
 channel gains from source $S$ and jammer $J$ to the reference antenna (the first
 antenna) of the { PSA} at destination $D$, respectively. It is worth pointing out that the
 eavesdropper $E$ in our work is a legitimate, active but non-intended receiver, which
 means that $E$ can simultaneously transmit signals to other nodes and intercept the
 confidential signal from source. Based on this assumption, the CSI of eavesdropper $E$
 is available through a training-based channel estimation technique. For the sake of
 improving security performance of the SIMO network, a beamforming vector
 $\bm{\omega}_d=\big[\omega_0 ~ \omega_1\cdots \omega_{N_D-1}\big]^{\rm T}$ satisfying
 $\big\|\bm{\omega}_d\big\|^2=1$ is applied to the $N_D$ antennas of the { PSA} to maximize
 the received confidential signal to interference plus noise ratio (SINR). Based on this
 setting, source $S$ and jammer $J$ simultaneously transmit the confidential signal
 $\widehat{s}_d$ and the jammer signal $\widehat{s}_j$ to the destination $D$ and
 eavesdropper $E$, respectively. Here, $\textsf{E}\big\{\big|\widehat{s}_d\big|\big\}^2=
 \textsf{E}\big\{\big|\widehat{s}_j\big|\big\}^2=1$ is assumed. Since $\widehat{s}_d$
 and $\widehat{s}_j$ are far field signals relative to the { PSA}, the signals $s_d$ and
 $s_j$ impinging on the reference antenna of the { PSA} from source $S$ and jammer $J$ are
 represented as $s_d=h_{SD}\sqrt{P_S}\widehat{s}_d$ and $s_j=h_{JD}\sqrt{P_J}\widehat{s}_j$,
 respectively, where $P_S$ and $P_J$ denote the maximum transmit powers of source $S$
 and jammer $J$, respectively.

\begin{figure*}[tp]\setcounter{equation}{26}
\vspace*{-1mm}
\begin{align}\label{f122} 
\begin{array}{cl}
 \min\limits_{P_S,P_J,\bm{p}} & P_S+P_J , \\
 \text{s.t.} & \frac{1+\sigma^{-2}\bm{p}^{\rm H}\big(P_S\big| h_{SD}\big|^2\bm{Q}_d^{\rm H}\bm{Q}_d
  +P_J\big| h_{JD}\big|^2\bm{Q}_j^{\rm H}\bm{Q}_j\big)\bm{p}+\sigma^{-4}P_S P_J
  \big| h_{SD}\big|^2\big| h_{JD}\big|^2 C_p}
  {1+\sigma^{-2}P_J\big| h_{JD}\big|^2\|\bm{a}_j\|^2} , \\
 & ~~~ \ge 2^{R_{\rm sec}^0}\big({1+P_S\bm{h}_{SE}^{\rm H}\big(P_J\bm{h}_{JE}\bm{h}_{JE}^{\rm H}
  + \sigma_e^2\bm{I}_{N_E}\big)^{-1} \bm{h}_{SE}}\big) , \\
 & \text{tr}\big(\bm{p}^{\rm T}\bm{F}_n\bm{p}\big)=1, \, 0\le n\le N_D -1, \ \ P_S\ge 0 , \, P_J \ge 0 .
\end{array}
\end{align}
\hrulefill
\vspace*{-4mm}
\end{figure*}

 Because both source $S$ and jammer $J$ are far-field narrowband { synchronized}
 transmitters{ \footnote{{Synchronizing the transmissions of source and jammer
 is important. To achieve the synchronization between two transmitters, one of the
 transmitters can serve as master and the other as slave, see for example \cite{syn}.
 In our case, the source serves as the master, who broadcasts the carrier and timing
 signals, while the jammer acts as the slave, who locks up to the carrier and timing
 signals from the master. In this way, the jammer acquires the carrier frequency and
 phase as well as achieves the timing synchronization with the source.}}},
 the change of the complex envelope of the corresponding EM signal when sweeping across the
 { PSA} is negligible. Therefore, the output signals at the { PSA} and the eavesdropper $E$ are
 given respectively as\setcounter{equation}{13}
\begin{align}
 y_D =& \bm{\omega}_d^{\rm H} \bm{Q}_d\bm{p} h_{SD} \sqrt{P_S}\widehat{s}_d +
  \bm{\omega}_d^{\rm H} \bm{Q}_j\bm{p} h_{JD} \sqrt{P_J}\widehat{s}_j+
  \bm{\omega}_d^{\rm H}\bm{n}_D , \label{ff5} \\
 y_E =&\bm{\omega}_e^{\rm H}\bm{ h}_{SE}\sqrt{P_S}\widehat{s}_d +
 \bm{\omega}_e^{\rm H} \bm{h}_{JE}\sqrt{P_J}\widehat{s}_j + \bm{\omega}_e^{\rm H}\bm{n}_E , \label{ff6}
\end{align}
 where $\bm{\omega}_e\in \mathbb{C}^{N_E}$ with $\big\|\bm{\omega}_e\big\|^2=1$ is the
 receive beamforming vector of eavesdropper $E$, while $\bm{n}_D \sim
 \mathcal{CN}\big(\bm{0},\sigma^2\bm{I}_{N_D}\big)$ and $\bm{n_E}\sim
 \mathcal{CN}\big(\bm{0},\sigma_e^2\bm{I}_{N_E}\big)$ are the received Gaussian noise
 vectors at destination $D$ and eavesdropper $E$, respectively. From the perspective of
 eavesdropper $E$, the optimal $\bm{\omega}_e$ is designed to achieve the maximum amount
 of wiretapped information, i.e., to maximize its desired SINR, which is obtained
 by solving the following problem
\begin{align}\label{eq16}
 & \max\limits_{\bm{\omega}_e} \frac{\bm{\omega}_e^{\rm H}\bm{h}_{SE}\bm{h}_{SE}^{\rm H}\bm{\omega}_e}
  {\bm{\omega}_e^{\rm H}\big(P_J\bm{h}_{JE}\bm{h}_{JE}^{\rm H}+\sigma_e^2\bm{I}_{N_E}\big)\bm{\omega}_e} .
\end{align}
 Clearly, the above problem is a standard generalized Rayleigh quotient problem, whose
 optimal solution is the generalized eigenvector corresponding to the largest generalized
 eigenvalue of the matrix pencil $\big(\bm{h}_{SE}\bm{h}_{SE}^{\rm H},P_J\bm{h}_{JE}
 \bm{h}_{JE}^{\rm H}+\sigma_e^2\bm{I}_{N_E}\big)$ \cite{GENRE}. Owing to the fact that
 the matrix $P_J\bm{h}_{JE}\bm{h}_{JE}^{\rm H}+\sigma_e^2\bm{I}_{N_E}$ is nonsingular,
 the optimal eavesdropper's receive beamforming vector $\bm{\omega}_e$ is equivalent to
 the normalized eigenvector associated with the maximum eigenvalue of the matrix
 $\big(P_J\bm{h}_{JE}\bm{h}_{JE}^{\rm H}+\sigma_e^2\bm{I}_{N_E}\big)^{-1}\bm{h}_{SE}
 \bm{h}_{SE}^{\rm H}$, that is,
\begin{align}\label{eq17}
 \bm{\omega}_e^{\star} =& c_e\bm{\vartheta}_{\max}\left(\big(P_J\bm{h}_{JE}\bm{h}_{JE}^{\rm H} +
  \sigma_e^2\bm{I}_{N_E}\big)^{-1}\bm{h}_{SE}\bm{h}_{SE}^{\rm H}\right) ,
\end{align}
 where $c_e$ is a normalized factor to satisfy $\Vert\bm{\omega}_e\Vert=1$ and
 $\bm{\vartheta}_{\max}(\mathbf{A})$ denotes the eigenvector corresponding to the
 maximum eigenvalue of the matrix $\mathbf{A}$. Considering the rank-1 property of
 the matrix $P_J\bm{h}_{JE}\bm{h}_{JE}^{\rm H}+\sigma_e^2\bm{I}_{N_E}$, the matrix
 $\big(P_J\bm{h}_{JE}\bm{h}_{JE}^{\rm H}+\sigma_e^2\bm{I}_{N_E}\big)^{-1}\bm{h}_{SE}
 \bm{h}_{SE}^{\rm H}$ is also rank-1 and only has one nonzero eigenvalue. Specifically,
 we have the formulation \eqref{eqr1} given at the top of this page. Thus the unique
 nonzero eigenvalue and  the corresponding eigenvector are $\bm{h}_{SE}^{\rm H}
 \big(P_J\bm{h}_{JE}\bm{h}_{JE}^{\rm H}+\sigma_e^2\bm{I}_{N_E}\big)^{-1}\bm{h}_{SE}$
 and $\big(P_J\bm{h}_{JE}\bm{h}_{JE}^{\rm H}+\sigma_e^2\bm{I}_{N_E}\big)^{-1}\bm{h}_{SE}$,
 respectively. Thus, the optimal eavesdropper's receive beamforming vector (\ref{eq17})
 can be written as\setcounter{equation}{18}
\begin{align}\label{eq19}
 \bm{\omega}_e^{\star} =& \frac{P_J(\bm{h}_{JE}\bm{h}_{JE}^H+\sigma_e^2\bm{I}_{N_E})^{-1}\bm{h}_{SE} }
  {\Vert(\bm{h}_{JE}\bm{h}_{JE}^H+\sigma_e^2\bm{I}_{N_E})^{-1}\bm{h}_{SE}\Vert} .
\end{align}

 Based on (\ref{ff5}) as well as (\ref{ff6}) and (\ref{eq19}), we formulate the received
 SINRs at destination $D$ and eavesdropper $E$ as
\begin{align} 
 \text{SINR}_D =& \frac{P_S\vert h_{SD}\vert^2\big|\bm{\omega}_d^{\rm H}\bm{Q}_d\bm{p}\big|^2}
 {\bm{\omega}_d^{\rm H}\big(\sigma^2\bm{I}_{N_D} + P_J\big| h_{JD}\big|^2\bm{Q}_j\bm{p} \bm{p}^{\rm T}
 \bm{Q}_j^{\rm H}\big)\bm{\omega}_d} , \label{f3} \\  
 \text{SINR}_E =& { \frac{P_S\big(\bm{\omega}_e^{\star}\big)^{\rm H}\bm{h}_{SE}\bm{h}_{SE}^{\rm H}
  \bm{\omega}_e^{\star}}{\big(\bm{\omega}_e^{\star}\big)^{\rm H}\big(P_J\bm{h}_{JE}\bm{h}_{JE}^{\rm H} +
  \sigma_e^2\bm{I}_{N_E}\big)\bm{\omega}_e^{\star}} } \nonumber \\
 =& P_S\bm{h}_{SE}^{\rm H}\big(P_J\bm{h}_{JE}\bm{h}_{JE}^{\rm H}+\sigma_e^2\bm{I}_{N_E}\big)^{-1}
  \bm{h}_{SE} , \label{f4} 
\end{align}
 respectively, where $\bm{p}^{\rm T}=\bm{p}^{\rm H}$ applies because $\bm{p}$ is a real
 vector. To realize secure communication of the SIMO network, the security metric called
 the maximum achievable secrecy rate \cite{wyner6} is considered, which is defined as follows
\begin{align}\label{f77} 
 R_{\rm sec} \le & \big[I(y_D,\widehat{s}_d)-I(y_E,\widehat{s}_d)\big]^{+} ,
\end{align}
 where $R_{\rm sec}$ denotes the achievable secrecy rate, $I(y_D,\widehat{s}_d)$ is the
 mutual information between source and destination, and $I(y_E,\widehat{s}_d)$ is the
 mutual information between source and eavesdropper. With the assumption of Gaussian
 wireless channels, $I(y_D,\widehat{s}_d)$ and $I(y_E,\widehat{s}_d)$ can readily be
 calculated as $I(y_D,\widehat{s}_d)=\log_2\big( 1 + \text{SINR}_D\big)$ and
 $I(y_E,\widehat{s}_d)=\log_2\big( 1 + \text{SINR}_E\big)$, respectively. Thus the
 maximum achievable secrecy rate of the SIMO network is formulated as
\begin{align}\label{f8} 
 R_{\text{sec}}^{\text{max}} \!\!=\!\!
  \left[ \log_2\frac{1 \!\!+\!\! \frac{P_S\big| h_{SD}\big|^2\big|\bm{\omega}_d^{\rm H}\bm{Q}_d\bm{p}\big|^2}
  {\bm{\omega}_d^{\rm H}\big(\sigma^2\bm{I}_{N_D} + P_J\big| h_{JD}\big|^2\bm{Q}_j\bm{p}
  \bm{p}^{\rm T}\bm{Q}_j^{\rm H}\big)\bm{\omega}_d}}{1\!\!+\!\! P_S\bm{h}_{SE}^{\rm H}\big(P_J\bm{h}_{JE}\bm{h}_{JE}^{\rm H}
  +\sigma_e^2\bm{I}_{N_E}\big)^{-1}\bm{h}_{SE}} \right]^{+} \!\! .
\end{align}
 For the point-to-point SIMO network, we consider two optimization problems, which are
 the total power minimization under secrecy rate constraint and the secrecy rate
 maximization under transmit power constraints, respectively.

\subsection{Total Power Minimization}\label{S3.1}

 The optimization problem is defined as the one that minimizes the total transmit power
 of the SIMO network subject to the minimum secrecy rate constraint $R_{\rm sec}^0$, that is,
\begin{equation}\label{f11} 
 \!\!\! \begin{array}{l}
 \min\limits_{P_S,P_J,\bm{p},\bm{\omega}_d} ~~ P_S+P_J ,\\
 \text{s.t. } \log_2\frac{1 + \frac{P_S\big| h_{SD}\big|^2\big|\bm{\omega}_d^{\rm H}\bm{Q}_d\bm{p}\big|^2}
  {\bm{\omega}_d^{\rm H}\big(\sigma^2\bm{I}_{N_D}+P_J\big| h_{JD}\big|^2\bm{Q}_j\bm{p}
  \bm{p}^{\rm T}\bm{Q}_j^{\rm H}\big) \bm{\omega}_d}} {1+P_S\bm{h}_{SE}^{\rm H}\big(P_J\bm{h}_{JE}\bm{h}_{JE}^{\rm H}
  +\sigma_e^2\bm{I}_{N_E}\big)^{-1} \bm{h}_{SE}} \ge R_{\rm sec}^0 , \\
  ~~~ P_S\ge 0 , P_J\ge 0 , \text{tr}\big(\bm{p}^{\rm T}\bm{F}_n\bm{p}\big)=1 , 0\le n\le N_D-1 ,
\end{array} \!\!\!
\end{equation}
 where $\bm{F}_n=\big(\bm{F}_n^{\frac{1}{2}}\big)^{\rm T}\bm{F}_n^{\frac{1}{2}}$ and
 $\bm{F}_n^{\frac{1}{2}}=\big[\bm{0}_{3\times n} ~ \bm{F} ~ \bm{0}_{3\times (N_D-n-1)}\big]$,
 in which the sparse matrix $\mathbf{F}\in\mathbb{R}^{3\times (2N_D+1)}$ is  defined as
\begin{equation}\label{eq25}
 \mathbf{F}[i,j] \! = \! \left\{ \!\!\! \begin{array}{l}
  1, \ (i,j)\! \in \! \{(1,1), (2,N_D\! +\! 1), (3,2N_D\! +\! 1)\} , \\
  0, \ \text{otherwise} . \end{array} \right. \!\!
\end{equation}
 Note that the  constraint $\text{tr}\big(\bm{p}^{\rm T}\bm{F}_n\bm{p}\big) =1$ in
 \eqref{f11} is equivalent to the property of { PSA} spatial pointings given in (\ref{eq7}).
 When $P_S$ and $P_J$ are given, the optimal $\bm{\omega}_d$ for the problem \eqref{f11}
 is obtained, similar to the derivation of $\bm{\omega}_e^{\star}$, as
\begin{align}\label{f5} 
 \bm{\omega}_d^{\rm opt} =& \frac{(\sigma^2\bm{I}_{N_D} + P_J\big| h_{JD}\big|^2\bm{Q}_j\bm{p}\bm{p}^{\rm T}
  \bm{Q}_j^{\rm H}\big)^{-1}\bm{Q}_d\bm{p}}{\Vert(\sigma^2\bm{I}_{N_D}
  +P_J\big| h_{JD}\big|^2\bm{Q}_j\bm{p}\bm{p}^{\rm T}
  \bm{Q}_j^{\rm H}\big)^{-1}\bm{Q}_d\bm{p} \Vert} .
\end{align}
 Next we substitute (\ref{f5}) into (\ref{f11}) to reformulate the total power
 minimization problem as \eqref{f122}, which is given at the top of this page.
 Unfortunately, because of the nonlinear and coupled term $C_p$, which is given by\setcounter{equation}{27}
\begin{align}\label{eq28}
 C_p \!\!=\!\! \bm{p}^{\rm T}\bm{Q}_d^{\rm H}\bm{Q}_d\bm{p}\bm{p}^{\rm T}\bm{Q}_j^{\rm H}\bm{Q}_j\bm{p}
 \!\!-\!\!\bm{p}^{\rm T}\bm{Q}_d^{\rm H}\bm{Q}_j\bm{p}\bm{p}^{\rm T}\bm{Q}_j^{\rm H}\bm{Q}_d\bm{p} \!\!\ge\!\! 0 ,
\end{align}
 the optimization problem (\ref{f122}) is generally nonconvex and difficult to solve
 directly. Hence we propose a suboptimal algorithm for the optimization problem
 {(\ref{f122}), i.e.,} (\ref{f11}). With this method, the optimization of $\bm{p}$ is
 performed independently from $P_S$ and $P_J$. Specifically, since the received
 desired signal strength at destination $D$ in the SIMO network satisfies
\begin{align}\label{eq29}
  P_S\big| h_{SD}\big|^2\big|\bm{\omega}_d^{\rm H}\bm{Q}_d\bm{p}\big|^2 \le &
  P_S\big| h_{SD}\big|^2\big\|\bm{\omega}_d\big\|^2\big\|\bm{Q}_d\bm{p}\big\|^2 \nonumber \\
  =& P_S\big| h_{SD}\big|^2\text{tr}\big(\bm{Q}_d^{\rm H}\bm{Q}_d\bm{p}\bm{p}^{\rm T}\big),
\end{align}
 we can consider the term $P_S\big| h_{SD}\big|^2\text{tr}\big(\bm{Q}_d^{\rm H}\bm{Q}_d
 \bm{P}_c\big)$ as the optimization objective for the { PSA} spatial pointings $\bm{p}$ by introducing
 $\bm{P}_c=\bm{p}\bm{p}^{\rm T}$. Thus, the secrecy optimization problem with respect to $\bm{p}$
 can be formulated as
\begin{equation}\label{optp} 
 \begin{array}{rl}
  \max\limits_{\bm{P}_c} & P_S\big| h_{SD}\big|^2\text{tr}\big(\bm{Q}_d^{\rm H}\bm{Q}_d\bm{P}_c\big) , \\
  \text{s.t.} & \bm{P}_c\succeq 0 , \, \text{rank}\big(\bm{P}_c\big)=1, \,
  \text{tr}\big(\bm{Q}_j^{\rm H}\bm{Q}_j\bm{P}_c\big)=0, \\
   &\text{tr}(\bm{F}_n\bm{P}_c\big )=1,\, 0\le n\le N_D-1.
 \end{array}
\end{equation}
 where the constraint $\text{tr}\big(\bm{Q}_j^{\rm H}\bm{Q}_j\bm{P}_c\big)=0$ indicates
 that the interference introduced by jammer $J$ to destination $D$ can be canceled
 completely. However, the problem \eqref{optp} is nonconvex and NP-hard due to the
 rank-1 constraint.

 In order to find an efficient way of solving the optimization \eqref{optp}, we firstly
 relax it to a standard semidefinite programming (SDP) problem by neglecting the rank-1
 constraint temporarily. Then the penalty based method \cite{12} is utilized to obtain
 the finally rank-1 satisfied solution for the problem \eqref{optp}. To be specific, let
 $\bm{P}_c^{\rm opt}$ be the optimal solution of (\ref{optp}) without considering the
 rank-1 constraint. Then $\text{tr}\big(\bm{Q}_d^{\rm H}\bm{Q}_d\bm{P}_c^{\rm opt}\big)$
 is actually an upper bound of $\text{tr}\big(\bm{Q}_d^{\rm H}\bm{Q}_d\bm{P}_c\big)$
 in the objective function of the problem \eqref{optp}. With the penalty based method,
 this $\bm{P}_c^{\rm opt}$ is adopted as the initial point $\bm{P}_c^{(0)}$ for the
 iterative optimization given in \eqref{optpp}:
\begin{equation}\label{optpp} 
 \!\!\! \begin{array}{rl}
 \bm{P}_c^{(t+1)} & \!\!\! = \arg  \min\limits_{\bm{P}_c}  \text{tr}\big(\bm{P}_c\big) \!\!-\!\!
  \lambda_{\max}\big(\bm{P}_c^{(t)}\big) \\
 & \hspace*{2mm} ~~-\text{tr}\left(\bm{\vartheta}_{\max}^{(t)}
  \big(\bm{\vartheta}_{\max}^{(t)}\big)^{\rm H}\big(\bm{P}_c\!\!-\!\!\bm{P}_c^{(t)}\big)\right) , \\
  \text{s.t.} & \!\!\! \text{tr}\big(\bm{Q}_d^{\rm H}\bm{Q}_d\bm{P}_c\big) \le \gamma, \bm{P}_c\succeq 0 ,
  \text{tr}\big(\bm{Q}_j^{\rm H}\bm{Q}_j\bm{P}_c\big)\! =\! 0, \\
 & \text{tr}\big(\bm{F}_n\bm{P}_c\big)\!=\!1, \,  0\le n\le N_D-1,
 \end{array}
\end{equation}
 where the auxiliary variable $\gamma$ satisfying $0\le \gamma\le
 \text{tr}\big(\bm{Q}_d^{\rm H}\bm{Q}_d\bm{P}_c^{\rm opt}\big)$, and the superscript
 $^{(t)}$ denotes the iteration number, while $\lambda_{\max}\big(\bm{P}_c^{(t)}\big)$ is
 the maximum eigenvalue of $\bm{P}_c^{(t)}$ and $\bm{\vartheta}_{\max}^{(t)}$ denotes the
 corresponding eigenvector. For a fixed $\gamma$, we can obtain the optimal rank-1
 satisfied solution $\bm{P}_c^{\rm opt}$ by solving the optimization problem \eqref{optpp}
 iteratively, and the corresponding optimal $\bm{p}^{\rm opt}$ is calculated through the
 eigenvalue decomposition of $\bm{P}_c^{\rm opt}$. We utilize  the bisection method
 \cite{bisection} to perform one-dimensional search for obtaining the optimal auxiliary
 variable $\gamma^{\star}$, so as to obtain the optimal solution $\bm{p}^{\star}$.
 The convergence of utilizing this penalty based method to solve the problem \eqref{optpp}
 {is proved in Appendix.}

 Once the optimal $\bm{p}^{\star}$ is given, { the optimal $\bm{\omega}_d^{\star}$ and the
 SINR at destination are derived respectively from (\ref{f5}) and (\ref{f3}) as}
\begin{align} 
 \bm{\omega}_d^{\star} =& {\bm{Q}_d\bm{p}^{\star}}/{\Vert\bm{Q}_d\bm{p}^{\star}\Vert},
 \label{optww} \\
 \text{SINR}_D =& \sigma^{-2}P_S\vert h_{SD}\vert^2\Vert\bm{Q}_d\bm{p}^{\star}\Vert^2 . \label{eq33}
\end{align}
 By substituting $\bm{p}^{\star}$ and $\bm{\omega}_d^{\star}$ into the original problem
 (\ref{f11}),  the reformulated total power minimization problem is given by
\begin{equation}\label{optpower} 
 \!\!\!\!\begin{array}{cl}
  \min\limits_{P_S,P_J} & \!\!\! P_S+P_J , \\
  \text{s.t.} & \!\!\! \log_2\left( \frac{1+\sigma^{-2}P_S\vert h_{SD}\vert^2\Vert\bm{Q}_d\bm{p}^{\star}\Vert^2}
  {{1+P_S\bm{h}_{SE}^{\rm H}(P_J \bm{h}_{JE}\bm{h}_{JE}^{\rm H}+\sigma_e^2\bm{I}_{N_E})^{-1}
  \bm{h}_{SE}}}\right) \! \ge \! R_{\rm sec}^0 , \\
  & \!\!\! P_S\ge 0 , \, P_J\ge 0 .
 \end{array} \!\!
\end{equation}
 After performing some mathematical transformations, we have
\begin{equation}\label{GP} 
 \begin{array}{cl}
 \min\limits_{P_S,P_J} & \!\! P_S+P_J , \\
 \text{s.t.} & \!\! {\sigma_e^2(2^{R_{\rm sec}^0}-1)} \!+\! (2^{R_{\rm sec}^0}-1)\Vert \bm{h}_{JE}\Vert^2 P_J \\
  & \!\! +\!({2^{R_{\rm sec}^0}\Vert \bm{h}_{SE}\Vert^2\!\!-\!\!\sigma_e^2\sigma^{-2}\vert h_{SD}\vert^2
  \Vert \bm{Q}_d\bm{p}^{\star}\Vert^2})P_S \\
  & \!\! +\!\big(2^{R_{\rm sec}^0}\sigma_e^{-2}a \\ & \!\!
  -\sigma^{-2}\vert h_{SD}\vert^2 \Vert \bm{Q}_d\bm{p}^{\star}\Vert^2\Vert \bm{h}_{JE}\Vert^2\big)P_SP_J\le 0 , \\
 & \!\! P_S\ge 0 , \, P_J\ge 0 ,
 \end{array} \!\!
\end{equation}
 where $a= \Vert \bm{h}_{SE}\Vert^2 \Vert\bm{h}_{JE}\Vert^2 -\vert \bm{h}_{SE}^{\rm H}\bm{h}_{JE}\vert^2$.
 For effectively solving the optimization problem \eqref{GP}, we consider different cases
 of the required secrecy rate threshold $R_{\rm sec}^0$, which corresponds to different
 optimal solutions of $P_S+P_J$. Firstly, two bounds of $R_{\rm sec}^0$ are defined as
\begin{align}
 R_1 &= \log_2\left(\frac{\sigma^{-2}\vert h_{SD}\vert^2\Vert
  \bm{Q}_d\bm{p}^{\star}\Vert^2}{\sigma_e^{-2}\Vert \bm{h}_{SE}\Vert^2}\right) , \label{eq36} \\
 R_2 &= \log_2\left(\frac{\sigma^{-2}\vert h_{SD}\vert^2\Vert \bm{Q}_d\bm{p}^{\star}\Vert^2\Vert
  \bm{h}_{JE}\Vert^2}{\sigma_e^{-2}a}\right) , \label{lbo} 
\end{align}
 Based on (\ref{eq36}) and \eqref{lbo}, the following three cases of $R_{\rm sec}^0$
 are discussed.

\subsubsection{Case~1.~$R_1 < R_{\rm sec}^0 < R_2$}

 In this case, the optimization problem \eqref{GP} is actually a standard geometric
 programming (GP) problem, which is
\begin{equation}\label{GP2} 
 \begin{array}{cl}
 \min\limits_{P_S,P_J} & \!\! P_S+P_J , \\
 \text{s.t.} &   g_2P_S^{-1}+g_3P_J^{-1} + g_1 P_S^{-1}P_J^{-1}\le 1\\
 & \!\! P_S\ge 0 , \, P_J\ge 0 .
 \end{array}
\end{equation}
 where
\begin{align}
 g_1 =& \frac{\sigma_e^2(2^{R_{\rm sec}^0}-1)}{(\sigma^{-2}\vert h_{SD}\vert^2 \Vert
  \bm{Q}_d\bm{p}^{\star}\Vert^2\Vert \bm{h}_{JE}\Vert^2-2^{R_{\rm sec}^0}\sigma_e^{-2}a)} , \label{eq39} \\
 g_2 =& \frac{(2^{R_{\rm sec}^0}-1)\Vert \bm{h}_{JE}\Vert^2}{(\sigma^{-2}\vert h_{SD}\vert^2 \Vert
 \bm{Q}_d\bm{p}^{\star}\Vert^2\Vert \bm{h}_{JE}\Vert^2-2^{R_{\rm sec}^0}\sigma_e^{-2}a)}, \label{eq40} \\
 g_3 =& \frac{({2^{R_{\rm sec}^0}\Vert \bm{h}_{SE}\Vert^2-\sigma_e^2\sigma^{-2}\vert h_{SD}\vert^2
  \Vert \bm{Q}_d\bm{p}^{\star}\Vert^2})}{(\sigma^{-2}\vert h_{SD}\vert^2 \Vert
  \bm{Q}_d\bm{p}^{\star}\Vert^2\Vert \bm{h}_{JE}\Vert^2-2^{R_{\rm sec}^0}\sigma_e^{-2}a)} . \label{eq41}
\end{align}
 Obviously, this optimization can be  efficiently solved using the convex optimization
 technique to yield corresponding optimal total transmit power $P_S^{\star}+P_J^{\star}$.

\subsubsection{Case~2.~$R_{\rm sec}^0\le R_1$}

 In fact, the expression $R_1$ denotes the maximum secrecy rate of the SIMO network
 without introducing jammer $J$ under a high SINR condition. If $R_{\rm sec}^0\le R_1$
 is required, it makes no sense to introduce jammer $J$ and thus $P_J=0$ is designed.
 Therefore, the optimization problem \eqref{GP} is transformed into
\begin{align}\label{eq42}
 \!\! \begin{array}{cl}
 \min\limits_{P_S} &\!\! P_S , \\
 \text{s.t.} &\!\! \big({2^{R_{\rm sec}^0}\Vert \bm{h}_{SE}\Vert^2-\sigma_e^2\sigma^{-2}
  \vert h_{SD}\vert^2 \Vert \bm{Q}_d\bm{p}^{\star}\Vert^2}\big)P_S \\
  & +{\sigma_e^2\big(2^{R_{\rm sec}^0}-1\big)} \le 0 , \\
  & \!\! P_S\ge 0 ,
\end{array} \!\!
\end{align}
 which has the optimal source transmit power $P_S^{\star}$ as
\begin{align}\label{eq43}
 P_S^{\star} =& \frac{\sigma_e^2\big(2^{R_{\rm sec}^0}-1\big)}
  {{2^{R_{\rm sec}^0}\Vert \bm{h}_{SE}\Vert^2-\sigma_e^2\sigma^{-2}\vert h_{SD}\vert^2
  \Vert \bm{Q}_d\bm{p}^{\star}\Vert^2}} .
 \end{align}
 In this case, the optimal total power consumption is then given by $P_S^{\star}+P_J^{\star}=P_S^{\star}$.

\subsubsection{Case~3.~$R_{\rm sec}^0\ge R_2$}

 When jammer $J$ is introduced to promote secure communications of the SIMO network, the
 maximum achievable secrecy rate is expressed as $R_2$. That is, if the required secrecy rate
 threshold $R_{\rm sec}^0\ge R_2$, the optimization problem \eqref{GP} is infeasible.

 In summary, by combining the optimization problems (\ref{optpp}), (\ref{optww}) and (\ref{GP}),
 the total power minimization problem \eqref{f11} can be solved efficiently in a suboptimal
 way by optimizing the {PSA} spatial pointings $\bm{p}$, the receive beamforming vector
 $\bm{\omega}_d$ and the total transmit power $P_S+P_J$, separately.

\subsection{Secrecy Rate Maximization}\label{S3.2}

 We now investigate the secrecy rate maximization of the SIMO network subject to the total
 transmit power constraint $P_{\max}$. Similar to the total power minimization of
 \eqref{f11}, the secrecy rate optimization problem is formulated as
\begin{equation}\label{f19} 
 \begin{array}{cl}
 \max\limits_{P_S,P_J,\bm{\omega}_d,\bm{p}} &
  \log_2 \frac{ 1 + \frac{P_S\vert h_{SD}\vert^2\vert\bm{\omega}_d^{\rm H}\bm{Q}_d\bm{p}\vert^2}
  {\bm{\omega}_d^{\rm H}\big(\sigma^2\bm{I}_{N_D}+P_J\vert h_{JD}\vert^2\bm{Q}_j\bm{p}
  \bm{p}^{\rm T}\bm{Q}_j^{\rm H}\big)\bm{\omega}_d} }
  { {1+P_S\bm{h}_{SE}^{\rm H}(P_J \bm{h}_{JE}\bm{h}_{JE}^{\rm H}+ \sigma_e^2\bm{I}_{N_E})^{-1} \bm{h}_{SE}} } , \\
 \text{s.t.} & \text{tr}\big(\bm{p}^{\rm T}\bm{F}_n\bm{p}\big)=1, \, 0\le n\le N_D-1 , \\
 & P_S+P_J\le P_{\max} , \, P_J \ge 0 , \, P_S\ge 0 .
 \end{array}
\end{equation}
{ Likewise, the problem \eqref{f19} is difficult to solve directly. However, it is known that  the secrecy rate maximization problem with total power constraint is equivalent to  the  total
 power  minimization problem with the  secrecy rate threshold in essence. As a result,  we can also apply the proposed suboptimal
algorithm for  problem \eqref{f11} to  the problem \eqref{f19}. The concrete solutions are presented   as follows.
Firstly, according to \eqref{optp}, the PSA spatial pointings $\bm{P}$ is optimized to maximize the   received signal strength  at destination with eliminating     the  interference introduced by jammer.  Then the joint optimization among  remaining variables $\{P_S,P_J,\bm{\omega}_d\}$ is performed.  Particularly, the optimal $\bm{\omega}_d$ for problem \eqref{f19} is not related to $\{P_S,P_J\}$ due to the zero interference  $\text{tr}\big(\bm{Q}_j^{\rm H}\bm{Q}_j\bm{P}_c\big)=0$ required in \eqref{optp}. As such, we  can derive the optimal $\bm{\omega}_d$ by maximizing the destination  SINR as in \eqref{optww}.  Further,  the joint optimization of $\{P_S,P_J\}$ is presented in the following  problem \eqref{f1999} based on the obtained  $ \bm{P}$ and $\bm{\omega}_d$. }

\begin{equation}\label{f1999} 
 \begin{array}{cl}
 \max\limits_{P_S,P_J} & \log_2 \frac{1+\sigma^{-2}P_S\vert h_{SD}\vert^2\Vert\bm{Q}_d\bm{p}^{\star}\Vert^2}
 {1+P_S\bm{h}_{SE}^{\rm H}(P_J \bm{h}_{JE}\bm{h}_{JE}^{\rm H}+ \sigma_e^2\bm{I}_{N_E})^{-1} \bm{h}_{SE}} , \\
 \text{s.t.} & P_S+P_J\le P_{\max} , \,  P_J \ge 0 , P_S\ge 0 .
 \end{array}
\end{equation}
 It is natural that the optimal solution of the problem (\ref{f1999}) is achieved when the
 constraint $P_S+P_J = P_{\max}$ holds. Therefore, we further rewrite the problem (\ref{f1999}) as
\begin{align}\label{f211} 
 \max\limits_{0\le P_S \le P_{\max}} f(P_S)  ,
\end{align}
 where
\begin{align}\label{eq47}
 f(P_S) = \frac{l_5P_S^2-l_4P_S-l_1}{l_3P_S^2-l_2P_S-l_1},
\end{align}
 and
\begin{equation}\label{eq48}
 \left\{ \begin{array}{l}
 l_1=\sigma_e^2+P_{\max}\Vert \bm{h}_{JE}\Vert^2 , \\
 l_2=\Vert \bm{h}_{SE}\Vert^2-\Vert \bm{h}_{JE}\Vert^2 +P_{\max} l_3, \\ l_3=\sigma_e^{-2}a \\
 l_4=\sigma^{-2}\vert h_{SD}\vert^2  \Vert\bm{Q}_d\bm{p}^{\star}\Vert^2\big(P_{\max}\vert
  \bm{h}_{JE}\vert^2+\sigma_e^2\big) -\Vert \bm{h}_{JE}\Vert^2 , \\
 l_5=\sigma^{-2}\Vert\bm{Q}_d\bm{p}^{\star}\Vert^2\vert h_{SD}\vert^2\Vert \bm{h}_{JE}\Vert^2 .
 \end{array} \right.
\end{equation}
 It can be seen that the optimization problem \eqref{f211} is an unconstrained quadratically
 fractional function maximization problem, whose the optimal solution $P_S^{\star}$ can
 easily be derived by the quadratic discriminant method, which is
\begin{align}\label{eq49}
 P_S^{\star} =& \min \big\{ P_{\max},P_S^{'}\big\} ,
\end{align}
 with
\begin{align}\label{eq50}
 P_S^{'} \! =& \! \left[\frac{-l_1(l_3\! -\! l_5)\! +\!
  \sqrt{l_1^2(l_3\! -\! l_5)^2\! -\! l_1(l_3l_4\! -\! l_5l_2)(l_4\! -\! l_2)}}{l_3l_4-l_5l_2}\right]^{+} \!\! .
\end{align}
 Once the optimal $P_S^{\star}$ is obtained, the optimal $P_J^{\star}=P_{\max}-P_S^{\star}$.
 Given the optimal source and jammer transmit powers $P_S^{\star}$ and $P_J^{\star}$, we can
 accordingly determine the maximum achievable secrecy rate of the SIMO network via \eqref{f8}.
{It is worth re-iterating that the `optimal' PSA spatial pointing vector
 $\bm{p}^{\star}$, the receive beamforming vector $\bm{\omega}_d^{\star}$, the transmit power
 pairs $P_S^{\star}$ and $P_J^{\star}$ so obtained do not offer an optimal solution of the
 optimization problem (\ref{f19}). Rather they only provide a suboptimal solution.}

\section{Relay Aided Secrecy Beamforming Design}\label{S4}

 We now extend the secure beamforming design to the relaying network with { PSA}.
 Specifically, two cases are considered, where the first case assumes that the perfect CSI
 in the relay network is available and the other one considers the imperfect {PSA} spatial
 pointings. For the both cases, we aim at improving the secrecy rate of the relaying network
 as much as possible.

\begin{figure}[htb]
\vspace*{-1mm}
\begin{center}
\includegraphics[width=0.50\columnwidth,height=0.42\columnwidth]{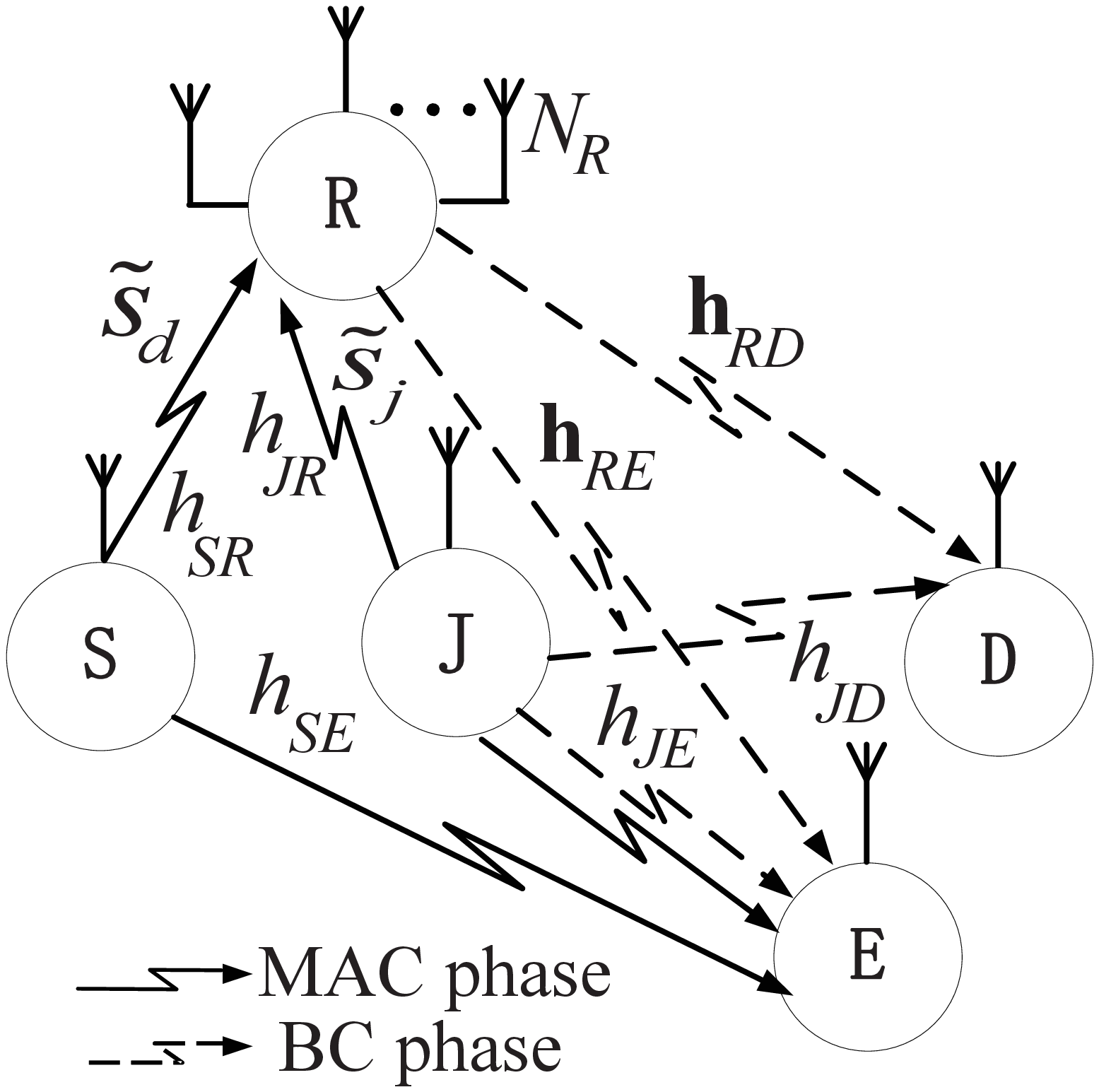}
\includegraphics[width=0.48\columnwidth,height=0.40\columnwidth]{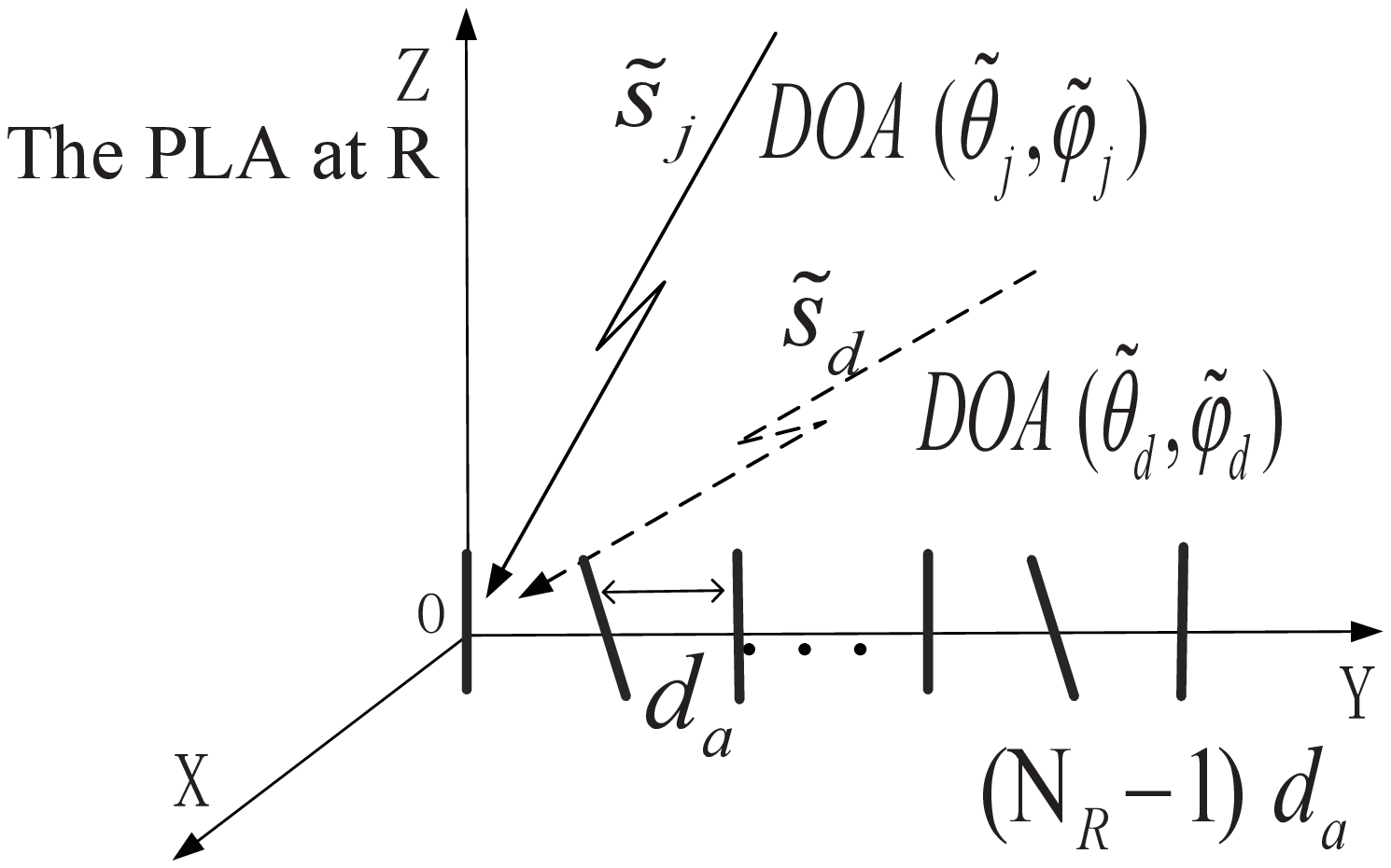}
\end{center}
\vspace*{-5mm}
\caption{A two-hop network with the polarization sensitive array having $N_R$ antennas at relay.}
\label{FIG3}
\vspace*{-1mm}
\end{figure}

 As shown in Fig.~\ref{FIG3}, source $S$, destination $D$, eavesdropper $E$ and jammer
 $J$ are all equipped with single-antenna, while relay $R$ employs the $N_R$-antenna { PSA}.
 Owing to the limit coverage of $S$, there exists no direct communication link between
 $S$ and $D$. Therefore, $S$ transmits confidential signal $\widehat{\widetilde{s}}_d$ to
 $D$ via $R$. Specifically, in the first phase known as multiple access (MAC) phase, $S$
 transmits $\widehat{\widetilde{s}}_d$ to $R$, and then in the second phase called
 broadcast (BC) phase, $R$ forwards the received signal in the first phase to $D$. Again,
 owing to the existence of eavesdropper $E$, jammer $J$ is introduced to transmit jamming
 signal $\widehat{\widetilde{s}}_j$ to decrease the information leakage which happens
 between $S$ and $R$ as well as between $R$ and $D$. { The transmissions of source $S$ and
 jammer $J$ in the MAC phase are synchronized, while the transmissions of relay $R$ and
 jammer $J$ in the BC phase are also synchronized.} For this relay network, it is
 reasonable to assume that source transmits signal using maximum transmit power.

\begin{figure*}[tp]\setcounter{equation}{58}
\vspace*{-1mm}
\begin{align}
 \bm{O}_E =& \left[\begin{array}{cc}
 \sigma_e^2+P_J^{(1)}\vert h_{JE}\vert^2 & P_J^{(1)} h_{JE} h_{JR}^{*}\big(\bm{h}_{RE}
  \bm{R}_{\text{cor}}^{\frac{1}{2}}\bm{W}\widetilde{\bm{Q}}_j\widetilde{\bm{p}}\big)^{*} \\
  P_J^{(1)}\big(\bm{h}_{RE}\bm{R}_{\text{cor}}^{\frac{1}{2}}\bm{W}\widetilde{\bm{Q}}_j
 \widetilde{\bm{p}}\big) h_{JR} h_{JE}^{*} & P_J^{(1)}\vert h_{JR}\vert^2\vert \bm{h}_{RE}
  \bm{R}_{\text{cor}}^{\frac{1}{2}}\bm{W}\widetilde{\bm{Q}}_j\widetilde{\bm{p}}\vert^2 +
  K_e \end{array}\right] , \label{eq59} \\
 O_D =& P_J^{(1)}\vert h_{JR}\vert^2\vert\bm{h}_{RD}\bm{R}_{\text{cor}}^{\frac{1}{2}}
  \bm{W}\widetilde{\bm{Q}}_j\widetilde{\bm{p}}\vert^2 + P_J^{(2)}\vert h_{JD}\vert^2 +
  \sigma_r^2\Vert\bm{h}_{RD}\bm{R}_{\text{cor}}^{\frac{1}{2}}\bm{W}\Vert^2
  + \sigma_d^2 , \label{eq60}
\end{align}
\hrulefill
\vspace*{-4mm}
\end{figure*}

\subsection{Secrecy Rate Maximization with Perfect CSI}\label{S4.1}

 Similar to Section~\ref{S3}, the scalar $h_{mn}\sim {\cal CN}(0,\sigma_h^2)$ denotes
 the flat-fading and quasi-static channel from node $m$ to node $n$  where $m=S,J$ and
 $n=E$, while $h_{SR}$ and $h_{JR}$ denote the channel gains from source $S$ and jammer
 $J$ to the reference antenna of the relay's PSA, respectively. Furthermore, the CSI of
 eavesdropper $E$ is assumed to be available. Based on these assumptions, the received
 signals at the $N_R$ antennas of relay $R$ in the MAC phase can be expressed as\setcounter{equation}{50}
\begin{align}\label{fr5} 
 \bm{y}_R
 =& \widetilde{\bm{Q}}_d\widetilde{\bm{p}} h_{SR}\sqrt{P_S}\widehat{\widetilde{s}}_d +
 \widetilde{\bm{Q}}_j\widetilde{\bm{p}} h_{JR}\sqrt{P_J^{(1)}}\widehat{\widetilde{s}}_j +
 \bm{n}_R ,
\end{align}
 where $\widehat{\widetilde{s}}_d$ and $\widehat{\widetilde{s}}_j$  are the transmit
 signals of source $S$ and jammer $J$, respectively, with $\textsf{E}\big\{\vert\widehat{\widetilde{s}}_d
 \vert^2\big\}=\textsf{E}\big\{\vert\widehat{\widetilde{s}}_j\vert^2\big\}=1$,
 $P_S$ and $P_J^{(1)}$ are the transmit powers of source $S$ and jammer $J$, respectively,
 while $\bm{n}_R\in\mathbb{C}^{N_R}$ is the Gaussian noise vector at relay $R$ whose
 elements follow the distribution ${\cal{CN}}(0,\sigma_r^2)$. The spatio-polarized
 manifold matrices $\widetilde{\bm{Q}}_k\in\mathbb{C}^{N_R\times 3N_R}$ for $k=d,j$ are
 defined similarly to (\ref{eq13}), and the relay's {PSA} spatial pointing vector
 $\widetilde{\bm{p}} \in \mathbb{R}^{3N_R}$ is defined similarly to \eqref{eq12}. The
 wiretapped signal at $E$ in this phase is given by
\begin{align}\label{fe1} 
 y_E ^{(1)} =& h_{SE}\sqrt{P_S}\widehat{\widetilde{s}}_d + h_{JE}\sqrt{P_J^{(1)}}\widehat{\widetilde{s}}_j
  + n_E^{(1)} \nonumber \\ =& h_{SE}\sqrt{P_S}\widehat{\widetilde{s}}_d + \widehat{n}_E^{(1)} ,
\end{align}
 where the additive Gaussian noise $n_E^{(1)}$ follows the distribution ${\cal{CN}}(0,\sigma_e^2)$,
 and $\widehat{n}_E^{(1)}=h_{JE}\sqrt{P_J^{(1)}}\widehat{\widetilde{s}}_j+n_E^{(1)}$.

 In the BC phase, relay $R$ utilizes the amplify-and-forward (AF) strategy to forward the
 received signal $\bm{y}_R$. To be specific, the retransmitted signal is $\bm{y}_R^{'}=
 \bm{W}\bm{y}_R$, where $\bm{W}\in\mathbb{C}^{N_R\times N_R}$ denotes the AF beamforming
 matrix. Thus, the transmit power of $R$ is given by
\begin{align}\label{eq53}
 P_R =& P_S\vert h_{SR}\vert^2\Vert\bm{W}\widetilde{\bm{Q}}_d\widetilde{\bm{p}}\Vert^2 +
  P_J^{(1)}\vert h_{JR}\vert^2\Vert\bm{W}\widetilde{\bm{Q}}_j\widetilde{\bm{p}}\Vert^2 \nonumber \\
 & + \sigma^2\text{tr}\big(\bm{W}\bm{W}^{\rm H}\big) .
\end{align}
 Simultaneously, jammer $J$ sends the interference signal $\widehat{\widetilde{s}}_j^{(2)}$
 with power $P_J^{(2)}$ to $D$. Let $\bm{h}_{RD}\in\mathbb{C}^{1\times N_R}$ and $\bm{h}_{RE}
 \in\mathbb{C}^{1\times N_R}$ be the channel gain vectors from the $N_R$ antennas of relay
 $R$ to destination $D$ and eavesdropper $E$, respectively, while $h_{JD}$ denotes the
 channel gain from $J$ to $D$. Then the received signals at $D$ and $E$ are formulated
 respectively as
\begin{align}
 y_D =& \bm{h}_{RD}\bm{R}_{\text{cor}}^{\frac{1}{2}}\bm{y}_R^{'} +
  \sqrt{P_J^{(2)}}h_{JD}\widehat{\widetilde{s}}_j^{(2)} + {n}_D \nonumber\\
  =& \bm{h}_{RD}\bm{R}_{\text{cor}}^{\frac{1}{2}}\bm{W}\widetilde{\bm{Q}}_d\widetilde{\bm{p}}
  h_{SR}\sqrt{P_S}\widehat{\widetilde{s}}_d + \widehat{n}_D , \label{eq54} \\
 y_E ^{(2)} =& \bm{h}_{RE}\bm{R}_{\text{cor}}^{\frac{1}{2}}\bm{y}_R^{'} +
  \sqrt{P_J^{(2)}}h_{JE}\widehat{\widetilde{s}}_j^{(2)} + {n}_E^{(2)}\nonumber\\
  =& \bm{h}_{RE}\bm{R}_{\text{cor}}^{\frac{1}{2}}\bm{W}\widetilde{\bm{Q}}_d\widetilde{\bm{p}}
  h_{SR}\sqrt{P_S}\widehat{\widetilde{s}}_d + \widehat{n}_E^{(2)} . \label{eq55}
\end{align}
 Due to the fact that relay $R$ adopts the { PSA} as the transmit array, its antenna
 correlation matrix $\bm{R}_{\text{cor}}\in\mathbb{C}^{N_R\times N_R}$ must be
 considered, whose elements follow the exponential model of $\bm{R}_{\text{cor}}[n,m]
 =p^{\vert n-m\vert}$ for $1\le n,m\le N_R$ with constant $p$ \cite{expo}.
 The additive Gaussian noises at destination $D$ and eavesdropper $E$ are
 $n_D\sim {\cal CN}(0,\sigma_d^2)$ and $n_E^{(2)}\sim {\cal CN}(0,\sigma_e^2)$,
 respectively, while the equivalent noise-plus-interference terms $\widehat{n}_D$ and
 $\widehat{n}_E^{(2)}$ are given by
\begin{align}
 \widehat{n}_D =& \bm{h}_{RD}\bm{R}_{\text{cor}}^{\frac{1}{2}}\bm{W}\widetilde{\bm{Q}}_j
  \widetilde{\bm{p}} h_{JR}\sqrt{P_J^{(1)}}\widehat{\widetilde{s}}_j + \sqrt{P_J^{(2)}} h_{JD}
  \widehat{\widetilde{s}}_j^{(2)} \nonumber\\
 & +\bm{h}_{RD}\bm{R}_{\text{cor}}^{\frac{1}{2}}
  \bm{W}\bm{n}_{R} + n_D , \label{eq56} \\
 \widehat{n}_E^{(2)} =& \bm{h}_{RE}\bm{R}_{\text{cor}}^{\frac{1}{2}}\bm{W}\widetilde{\bm{Q}}_j
  \widetilde{\bm{p}} h_{JR}\sqrt{P_J^{(1)}}\widehat{\widetilde{s}}_j + \sqrt{P_J^{(2)}} h_{JE}
  \widehat{\widetilde{s}}_j^{(2)} \nonumber\\
 & + \bm{h}_{RE}\bm{R}_{\text{cor}}^{\frac{1}{2}}\bm{W}
  \bm{n}_R + n_E^{(2)} . \label{eq57}
\end{align}

 Clearly, the total amount of information leakage to $E$ comes from both $S$ and $R$, as
 indicated in (\ref{fe1}) and (\ref{eq55}). Hence, the wiretapped information in the relay
 network is given by\setcounter{equation}{57}
\begin{align}\label{eq58}
 \bm{y}_E =& \left[\begin{array}{c} h_{SE} \\
  \bm{h}_{RE}\bm{R}_{\text{cor}}^{\frac{1}{2}}\bm{W}\widetilde{\bm{Q}}_d\widetilde{\bm{p}} h_{SR}
  \end{array}\right] \sqrt{P_S}\widehat{\widetilde{s}}_d + \left[\begin{array}{c}
  \widehat{n}_{E}^{(1)} \\ \widehat{n}_{E}^{(2)} \end{array}\right] \nonumber \\
  =& \bm{H}_E \sqrt{P_S}\widehat{\widetilde{s}}_d + \widehat{\bm{n}}_E .
\end{align}
 The covariance matrix $\bm{O}_E$ of $\widehat{\bm{n}}_E$ and $O_D\! =\! \text{E}
 \big\{\big|\widehat{n}_D\big|^2\big\}$ are given by \eqref{eq59} and \eqref{eq60}, respectively,
 at the top of the next page, in which $K_e\! =\! P_J^{(2)}\vert h_{JE}\vert^2 \! +\!
 \sigma_r^2\Vert \bm{h}_{RE}^{\rm T} \bm{R}_{\text{cor}}^{\frac{1}{2}}\bm{W}\Vert^2\! +\! \sigma_d^2$.
 Correspondingly, the achievable secrecy rate region of this relaying network is\setcounter{equation}{60}
\begin{align}\label{f777} 
 \widetilde{R}_{\text{sec}} \le \big[\widetilde{I}\big(y_D,\widehat{\widetilde{s}}_d\big) -
  \widetilde{I}\big(y_E,\widehat{\widetilde{s}}_d\big)\big]^+ ,
\end{align}
 in which the mutual information between source $S$ and destination $D$ and the mutual
 information between source $S$ and eavesdropper $E$ are given respectively by
\begin{align}
 & \hspace*{-2mm}\widetilde{I}\big(y_D,\widehat{\widetilde{s}}_d\big) \!= \! \log_2( 1  \!+ \!
  {P_S\vert h_{SR}\vert^2\vert\bm{h}_{RD}\bm{R}_{\text{cor}}^{\frac{1}{2}}
  \bm{W}\widetilde{\bm{Q}}_d\widetilde{\bm{p}}\vert^2}/ O_D) , \! \label{eq62} \\
 & \hspace*{-2mm}\widetilde{I}\big(y_E,\widehat{\widetilde{s}}_d\big) \! = \! \log_2 \det \left( \bm{I}_2  \!+ \!
  P_S\bm{H}_{E}\bm{H}_{E}^{\rm H}\bm{O}_E^{-1}\right) .\label{eq63}
\end{align}
 The optimization problem for the proposed secure beamforming design is formulated as
\begin{equation}\label{eq64}
 \begin{array}{cl}
 \max\limits_{P_J^{(1)},P_J^{(2)},\bm{W},\widetilde{\bm{p}}} & \!\!\!
 \log_2 \frac{1 + \frac{P_S\vert h_{SD}\vert^2\vert\bm{h}_{RD}\bm{R}_{\text{cor}}^{\frac{1}{2}}
 \bm{W}\widetilde{\bm{Q}}_d\widetilde{\bm{p}}\vert^2}{O_D}}{\det\big(\bm{I}_2 + P_S\bm{H}_{E}\bm{H}_{E}^{\rm H}
 \bm{O}_{E}^{-1}\big)} , \\
 \text{s.t.} & \!\!\! \text{tr}\big(\widetilde{\bm{p}}^{\rm T}\widetilde{\bm{F}}_n\widetilde{\bm{p}}\big) = 1 ,
 \, 0\le n\le N_R-1 , \\
 & \!\!\! P_S\vert h_{SR}\vert^2\Vert\bm{W}\widetilde{\bm{Q}}_d\widetilde{\bm{p}}\Vert^2
  + P_J^{(1)}\vert h_{JR}\vert^2\Vert \bm{W}\widetilde{\bm{Q}}_j\widetilde{\bm{p}}\Vert^2  \\
 & \!  + \sigma_r^2\text{tr}\big(\bm{W}\bm{W}^{\rm H}\big)\le \!P_R^{\rm max} , \\
 & \!\!\! 0\le P_J^{(1)}+P_J^{(2)}\le P_J^{\rm max} ,
 \end{array}
\end{equation}
 where the definition of $\widetilde{\bm{F}}_n\in\mathbb{R}^{3N_R\times3N_R}$ is similar
 to that of $\bm{F}_n$ given in Section~\ref{S3.1}, while $P_R^{\rm max}$ and $P_J^{\rm max}$
 are the maximum relay and jammer transmit powers, respectively. It can be observed that the
 objective function of this problem is a product of two correlated generalized Rayleigh
 quotients and is obviously nonconvex. Thus this optimization is difficult to solve directly.
 Since eavesdropper $E$ is a legitimate although not an intended receiver, we assume that the
 perfect CSI of $E$ is available. Then the following operations are performed.

 \emph{1)}~As the perfect CSI of $E$ is available, the beamforming matrix $\bm{W}$ is
 designed to satisfy $\bm{h}_{RE}\bm{R}_{\text{cor}}^{\frac{1}{2}}\bm{W}
 \widetilde{\bm{Q}}_d\widetilde{\bm{p}}=0$. Thus the information leakage from $R$ to $E$
 is canceled completely. With this beamforming matrix design, jammer $J$ does not need
 to transmit signal $\widehat{\widetilde{s}}_j^{(2)}$ to decrease the information leakage
 caused by $R$, which means that $P_J^{(2)}=0$.

 \emph{2)}~As destination $D$ is disturbed by the forwarded jammer signal
 $\widehat{\widetilde{s}}_j$ from the MAC phase, the beamforming matrix $\bm{W}$ should
 be designed to satisfy $\bm{h}_{RD}\bm{R}_{\text{cor}}^{\frac{1}{2}}\bm{W}
 \widetilde{\bm{Q}}_j\widetilde{\bm{p}}=0$ to eliminate the interference to $D$ caused
 by jammer $J$ completely.

 \emph{3)}~Since only jammer signal $\widehat{\widetilde{s}}_j$ is utilized to decrease
 the information leakage to $E$ in the MAC phase and $P_J^{(2)}=0$, we can set
 $P_J^{(1)}=P_J^{\rm max}$ to interfere eavesdropper maximally. Thus the power allocation
 for jammer $J$ is determined.

\begin{figure*}[tp!]\setcounter{equation}{77}
\vspace*{-1mm}
\begin{align}
 &\bm{\mu} = \Big(\big(\bm{G}_{je}^{{(l-1)}{\bot}}\big)^{\rm H}\big(P_R^{\rm max}\bm{G}_b+
  P_S\vert h_{SR}\vert^2\bm{R}_{d}^{(l-1)}+P_J^{\rm max}\vert h_{JR}\vert^2\bm{R}_j^{(l-1)}+\sigma_r^2\bm{I}_{N_R^2}\big)
  \bm{G}_{je}^{{(l-1)}{\bot}}\Big)^{-1} \big(\bm{G}_{je}^{{(l-1)}{\bot}}\big)^{\rm H}\bm{g}_d^{(l-1)} ,\label{eq78} \\
 & c\big(\widetilde{\bm{p}}^{(l-1)}\big) = \sqrt{P_R^{\rm max}\Big/\Big({\bm{\mu}^{\rm H}\big(
 \bm{G}_{je}^{{(l-1)}{\bot}}\big)^{\rm H}\big(P_S\vert h_{SR}\vert^2\bm{R}_d^{(l-1)}+P_J^{\rm max}\vert h_{JR}\vert^2
 \bm{R}_j^{(l-1)}+\sigma_r^2\bm{I}_{N_R^2}\big)\bm{G}_{je}^{{(l-1)}{\bot}}\bm{\mu}}\Big)} .\label{eq79}
\end{align}
\hrulefill
\vspace*{-4mm}
\end{figure*}

\begin{figure*}[bp]\setcounter{equation}{85}
\vspace*{-4mm}
\hrulefill
\begin{equation}\label{optpsub3} 
\begin{array}{l}
 \widetilde{\bm{P}}_c^{[t+1]} = \arg \min\limits_{\widetilde{\bm{P}}_c}
  \text{tr}\big(\widetilde{\bm{P}}_c\big) - \lambda_{\max}\big(\widetilde{\bm{P}}_c^{[t]}\big)
  - \text{tr}\left(\widetilde{\bm{\upsilon}}_{\max}^{[t]}\big(\widetilde{\bm{\upsilon}}_{\max}^{[t]}\big)^{\rm H}
  \big(\widetilde{\bm{P}}_c - \widetilde{\bm{P}}_c^{[t]}\big)\right) -
  \text{tr}\big(\widetilde{\bm{R}}_d^{(l)}\widetilde{\bm{P}}_c\big) , \\
 \mathrm{s.t.} ~ \text{tr}\big(\widetilde{\bm{R}}_d^{(l)}\widetilde{\bm{P}}_c\big)\le \widetilde{\gamma} , \
  \text{tr}\left(\! \big(P_S\vert h_{SR}\vert^2\widetilde{\bm{G}}_d^{(l)} \! + \!
  P_J^{\max}\vert h_{JR}\vert^2 \widetilde{\bm{G}}_j^{(l)}\big)\widetilde{\bm{P}}_c\right) \! \le \!
  P_R^{\max} \! - \! \sigma_r^2\text{tr}\big(\bm{W}^{(l)}\big(\bm{W}^{(l)}\big)^{\rm H}\big)  , \\
  ~~~~~ \widetilde{\bm{P}}_c\succeq \bm{0} , \
  \text{tr}\Big(\widehat{\widetilde{\bm{F}}}_n^{{(l)}}\widetilde{\bm{P}}_c\Big)=1 , \, 0\le n\le N_D-1,
\end{array}
\end{equation}
\vspace*{-1mm}
\end{figure*}

 With the operations \emph{1)} to \emph{3)}, the information leakage only occurs in the MAC
 phase, and the mutual information $\widetilde{I}\big(y_D,\widehat{\widetilde{s}}_d\big)$
 and $\widetilde{I}\big(y_E,\widehat{\widetilde{s}}_d\big)$ are simplified as\setcounter{equation}{64}
 \begin{align} 
 &\hspace*{-2mm}\widetilde{I}\big(y_D,\widehat{\widetilde{s}}_d\big) \!= \! \log_2(1 \!+ \!
  {P_S\vert h_{SR}\vert^2\vert\bm{h}_{RD}\bm{R}_{\text{cor}}^{\frac{1}{2}}
  \bm{W}\widetilde{\bm{Q}}_d\widetilde{\bm{p}}\vert^2}/ O_D ) , \! \label{eqq1}
\end{align}
\vspace*{-6mm}
\begin{align}
 &\hspace*{-6mm}\widetilde{I}\big(y_E,\widehat{\widetilde{s}}_d\big) \! = \!
 \log_2  \left(1+\frac{P_S\vert h_{SE}\vert^2}{\sigma_e^2+P_J^{\rm max}\vert h_{JE}\vert^2}\right).\label{eqq2}
\end{align}
Thus the secrecy rate maximization problem (\ref{eq64}) can be re-expressed as
\begin{equation}\label{subrelay} 
 \begin{array}{cl}
 \max\limits_{\bm{W},\widetilde{\bm{p}}} & \!\!\!
 \log_2 \frac{1+\frac{P_S\vert h_{SR}\vert^2\vert\bm{h}_{RD}\bm{R}_{\text{cor}}^{\frac{1}{2}}
 \bm{W}\widetilde{\bm{Q}}_d\widetilde{\bm{p}}\vert^2}{\sigma_d^2+\sigma_r^2\Vert\bm{h}_{RD}
 \bm{R}_{\text{cor}}^{\frac{1}{2}}\bm{W}\Vert^2}}
 {1+\frac{P_S\vert h_{SE}\vert^2}{\sigma_e^2+P_J^{\rm max}\vert h_{JE}\vert^2}} , \\
 \mathrm{s.t.} & \!\!\! \text{tr}\big(\widetilde{\bm{p}}^{\rm T}\widetilde{\bm{F}}_n\widetilde{\bm{p}}\big) = 1 ,
 \, 0\le n\le N_R-1 , \\
 & \!\!\! \bm{h}_{RD}\bm{R}_{\text{cor}}^{\frac{1}{2}}
 \bm{W}\widetilde{\bm{Q}}_j\widetilde{\bm{p}}=0 , \, \bm{h}_{RE}\bm{R}_{\text{cor}}^{\frac{1}{2}}\bm{W}
 \widetilde{\bm{Q}}_d\widetilde{\bm{p}}=0  , \\
 & \!\!\! P_S\vert h_{SR}\vert^2\Vert \bm{W}\widetilde{\bm{Q}}_d\widetilde{\bm{p}}\Vert^2 \! +\!
  P_J^{\rm max}\vert h_{JR}\vert^2 \Vert \bm{W}\widetilde{\bm{Q}}_j\widetilde{\bm{p}}\Vert^2 \\
 & \!\!\! ~~ +\sigma_r^2\text{tr}(\bm{W}\bm{W}^H)\le P_R^{\rm max}  .
\end{array}
\end{equation}
 Unfortunately, this problem is still neither convex nor concave with respect to $\bm{W}$
 and $\widetilde{\bm{p}}$. Similar to solving (\ref{f11}), we propose an iterative suboptimal
 algorithm to solve (\ref{subrelay}) effectively.

\subsubsection{Optimization of $\bm{W}$}

 When the { PSA} spatial pointing vector is fixed to $\widetilde{\bm{p}}=\widetilde{\bm{p}}^{(l-1)}$
 where $l$ is the outer iteration index and $\text{tr}\big(\big(\widetilde{\bm{p}}^{(l-1)}\big)^{\rm T}
 \widetilde{\bm{F}}_n\widetilde{\bm{p}}^{(l-1)}\big)=1$ for $0\le n\le N_R-1$, the problem
 (\ref{subrelay}) is transformed into
 \begin{align}\label{optw4} 
 \max\limits_{\bm{W}} &
  {1+\frac{P_S\vert h_{SR}\vert^2\vert\bm{h}_{RD}\bm{R}_{\text{cor}}^{\frac{1}{2}}
  \bm{W}\widetilde{\bm{Q}}_d\widetilde{\bm{p}}^{(l-1)}\vert^2}{\sigma_d^2+\sigma_r^2\Vert\bm{h}_{RD}
  \bm{R}_{\text{cor}}^{\frac{1}{2}}\bm{W}\Vert^2}} , \nonumber\\
 \mathrm{s.t.} & \bm{h}_{RD}\bm{R}_{\text{cor}}^{\frac{1}{2}}
  \bm{W}\widetilde{\bm{Q}}_j\widetilde{\bm{p}}^{(l-1)}=0 , \nonumber\\
  & \bm{h}_{RE}\bm{R}_{\text{cor}}^{\frac{1}{2}}\bm{W} \widetilde{\bm{Q}}_d\widetilde{\bm{p}}^{(l-1)}=0, \nonumber\\
  & P_S\vert h_{SR}\vert^2\Vert \bm{W}\widetilde{\bm{Q}}_d\widetilde{\bm{p}}^{(l-1)}\Vert^2 + P_J^{\rm max}\vert h_{JR}\vert^2 \Vert \bm{W}\widetilde{\bm{Q}}_j\widetilde{\bm{p}}^{(l-1)}\Vert^2 \nonumber\\
  & ~~ +\sigma_r^2\text{tr}(\bm{W}\bm{W}^H)\le P_R^{\rm max}.
\end{align} After some manipulations, \eqref{optw4} can be rewritten as
\begin{equation}\label{optw} 
 \begin{array}{cl}
 \max\limits_{\bm{\omega}_R} & 1 + \frac{P_S\vert h_{SR}\vert^2\bm{\omega}_R^{\rm H}\bm{G}_d^{(l-1)}\bm{\omega}_R}
  {\sigma_d^2 + \sigma_r^2\bm{\omega}_R^{\rm H}\bm{G}_{b}\bm{\omega}_R} , \\
 \mathrm{s.t.} & \big(\bm{g}_j^{(l-1)}\big)^{\rm H}\bm{\omega}_R=0 , \, \big(\bm{g}_e^{(l-1)}\big)^{\rm H}\bm{\omega}_R=0 , \\
 & \bm{\omega}_R^{\rm H}\big(P_S\vert h_{SR}\vert^2\bm{R}_d^{(l-1)}\! +\! P_J^{\rm max}\vert h_{JR}\vert^2\bm{R}_j^{(l-1)}
 \\ &~~ \! +\! \sigma_r^2\bm{I}_{N_R^2}\big)\bm{\omega}_R \le P_R^{\rm max} ,
\end{array}
\end{equation}
 where
\begin{align}
 & \bm{\omega}_R \! =\! \text{vec}(\bm{W}) \in \mathbb{C}^{N_R^2} , \label{eq70} \\
 & \bm{g}_k^{(l-1)} \! =\! \big(\widetilde{\bm{Q}}_k\widetilde{\bm{p}}^{(l-1)}\big)^{\ast}\!\otimes\!
  \big(\bm{h}_{RD}\bm{R}_{\text{cor}}^{\frac{1}{2}}\big)^{\rm H}\in \mathbb{C}^{N_R^2},
  k=d,j, \label{eq71} \\
 & \bm{g}_e^{(l-1)} \! =\! \big(\widetilde{\bm{Q}}_d\widetilde{\bm{p}}^{(l-1)}\big)^{\rm \ast}\!\otimes\!
  \big(\bm{h}_{RE}\bm{R}_{\text{cor}}^{\frac{1}{2}}\big)^{\rm H} \in \mathbb{C}^{N_R^2} , \label{eq72} \\
 & \bm{G}_d^{(l-1)} \! =\! \bm{g}_d^{(l-1)}\big(\bm{g}_d^{(l-1)}\big)^{\rm H}, \label{eq73} \\
 & \bm{G}_b \! =\! \big(\bm{I}_{N_R}\!\otimes\!\big(\bm{h}_{RD}\bm{R}_{\text{cor}}^{\frac{1}{2}}\big)^{\rm H}
  \big)\big(\bm{I}_{N_R}\otimes\big(\bm{h}_{RD}\bm{R}_{\text{cor}}^{\frac{1}{2}}\big)\big) , \label{eq74} \\
 & \bm{R}_k^{(l-1)} = \big(\big(\widetilde{\bm{Q}}_k\widetilde{\bm{p}}^{(l-1)}\big)^{\ast} \! \otimes\!
  \bm{I}_{N_R}\big) \big(\big(\widetilde{\bm{Q}}_k\widetilde{\bm{p}}^{(l-1)}\big)^{\rm T}\!\otimes \! \bm{I}_{N_R}\big),
  \nonumber \\ & ~~~~~~~~~~~~ k=d,j .\label{eq75}
\end{align}
 Let $\bm{G}_{je}^{{(l-1)}{\bot}}$ be the projection matrix onto the null space of
 $\bm{G}_{je}^{(l-1)}=\big[\bm{g}_j^{(l-1)} ~ \bm{g}_e^{(l-1)}\big]^{\rm H}$. Then the
 $N_R^2$-dimensional relay beamforming vector is denoted  as $\bm{\omega}_R=
 \bm{G}_{je}^{{(l-1)}{\bot}}\widetilde{\bm{\omega}}_R$, which transforms the original
 optimization variable $\bm{\omega}_R$ into $\widetilde{\bm{\omega}}_R$. Thus, the
 problem (\ref{optw}) can be rewritten as
\begin{equation}\label{eq76}
 \hspace*{-3mm}\begin{array}{l}
 \max\limits_{\widetilde{\bm{\omega}}_R} ~ 1+\frac{P_S\vert h_{SR}\vert^2\widetilde{\bm{\omega}}_R^{\rm H}
 \big(\bm{G}_{je}^{{(l-1)}{\bot}}\big)^{\rm H}\bm{G}_d^{(l-1)}\bm{G}_{je}^{{(l-1)}{\bot}}\widetilde{\bm{\omega}}_R}
  {\sigma_d^2+\sigma_r^2\widetilde{\bm{\omega}}_R^{\rm H}\big(\bm{G}_{je}^{{(l-1)}{\bot}}\big)^{\rm H}
  \bm{G}_b\bm{G}_{je}^{{(l-1)}{\bot}}\widetilde{\bm{\omega}}_R} , \\
 \mathrm{s.t.} ~~ \widetilde{\bm{\omega}}_R^{\rm H}\big(\bm{G}_{je}^{{(l-1)}{\bot}}\big)^{\rm H}
  \big(P_S\vert h_{SR}\vert^2\bm{R}_d^{(l-1)}\! +\! \\
 ~~~~~~ P_J^{\rm max}\vert h_{JR}\vert^2\bm{R}_j^{(l-1)}
  \! +\! \sigma_r^2\bm{I}_{N_R^2}\big)\bm{G}_{je}^{{(l-1)}{\bot}}\widetilde{\bm{\omega}}_R\! \le\! P_R^{\rm max} \! . \!
\end{array} \!
\end{equation}
 The problem in \eqref{eq76} is also a generalized Rayleigh quotient problem, which has the
 closed-form solution
\begin{equation}\label{eq77}
 \widetilde{\bm{\omega}}_R^{(l)} = c\big(\widetilde{\bm{p}}^{(l-1)}\big)\bm{\mu},
\end{equation}
 with $\bm{\mu}$ and  $c\big(\widetilde{\bm{p}}^{(l-1)}\big)$ given by \eqref{eq78} and \eqref{eq79},
 respectively, at the top of this page. Once the optimal $\widetilde{\bm{\omega}}_R^{(l)}$ is obtained,
 the optimal $\bm{W}^{(l)}$ can be derived based on $\bm{\omega}_R^{(l)}=\bm{G}_{je}^{{(l-1)}{\bot}}
 \widetilde{\bm{\omega}}_R^{(l)}$.

\begin{figure*}[tp]\setcounter{equation}{89}
\vspace*{-1mm}
\begin{align}\label{f24} 
\begin{array}{l}
 \max\limits_{\bm{W}_{\text{rb}}} ~ \min\limits_{\Delta\widetilde{\bm{p}}} ~ \log_2
  {\Big(1\! +\! \frac{P_S\vert h_{SR}\vert^2\big\vert\bm{h}_{RD}^{\rm T}\bm{R}_{\text{cor}}^{\frac{1}{2}}
  \bm{W}_{\text{rb}}\widetilde{\bm{Q}}_d\big(\widetilde{\bm{p}}^{\star}+\Delta\widetilde{\bm{p}}\big)\big\vert^2}
  {P_J^{(1)}\vert h_{JR}\vert^2\big\vert\bm{h}_{RD}^{\rm T}\bm{R}_{\text{cor}}^{\frac{1}{2}}
  \bm{W}_{\text{rb}}\widetilde{\bm{Q}}_j\big(\widetilde{\bm{p}}^{\star}+\Delta\widetilde{\bm{p}}\big)
  \big\vert^2+P_J^{(2)}\vert h_{JD}\vert^2+\sigma_r^2\big\Vert\bm{h}_{RD}^{\rm T}\bm{R}_{\text{cor}}^{\frac{1}{2}}
  \bm{W}_{\text{rb}}\big\Vert^2+\sigma_d^2}\Big)}-\log_2
  {\text{det}\big(\bm{I}_2\! +\! P_S\bm{H}_{E}\bm{H}_{E}^{\rm H}\bm{O}_{E}^{-1}\big)} , \\
 \mathrm{s.t.} ~ P_S\vert h_{SR}\vert^2\Vert \bm{W}_{\text{rb}}\widetilde{\bm{Q}}_d(\widetilde{\bm{p}}^{\star}\!+
  \!\Delta\widetilde{\bm{p}})\Vert^2 \!+\!P_J^{(1)}\vert h_{JR}\vert^2\Vert \bm{W}_{\text{rb}}
  \widetilde{\bm{Q}}_j(\widetilde{\bm{p}}^{\star}\!+\!\Delta{\widetilde{\bm{p}}})\Vert^2
  \!+\!\sigma_r^2\text{tr}\big(\bm{W}_{\text{rb}}\bm{W}_{\text{rb}}^{\rm H}\big)\! \le\! P_R^{\rm max}, \ \
 0\le P_J^{(1)}+P_J^{(2)} \le P_J^{\rm max} .
\end{array}
\end{align}
\hrulefill
\begin{align}\label{f25} 
\begin{array}{l}
 \max\limits_{\bm{W}_{\text{rb}}} ~ \bar{\gamma} , \\
 \mathrm{s.t.} ~ \frac{P_S\vert h_{SR}\vert^2\big(\widetilde{\bm{p}}^{\star}\!+\! \Delta\widetilde{\bm{p}}\big)^{\rm H}
  \widehat{\widetilde{\bm{R}}}_d\big(\widetilde{\bm{p}}^{\star}\!+\! \Delta\widetilde{\bm{p}}\big)}
  {P_J^{\rm max}\vert h_{JR}\vert^2\big(\widetilde{\bm{p}}^{\star}\! +\! \Delta\widetilde{\bm{p}}\big)^{\rm H}
  \widehat{\widetilde{\bm{R}}}_j\big(\widetilde{\bm{p}}^{\star}\! +\! \Delta\widetilde{\bm{p}}\big)\! +
  \! \sigma_r^2\Vert\bm{h}_{RD}^{\rm T}\bm{R}_{\text{cor}}^{\frac{1}{2}}\bm{W}_{\text{rb}}\Vert^2\! +\! \sigma_d^2}
  \ge 2^{\bar{\gamma}}\left(1\!+\!\frac{P_S\vert h_{SE}\vert^2}{\sigma_e^2\!+\!P_J^{\rm max}\vert h_{JE}\vert^2}\right)
  -1, ~ \forall\Delta\widetilde{\bm{p}} , \\
 ~~~~ P_S\vert h_{SR}\vert^2\Vert \bm{W}_{\text{rb}}\widetilde{\bm{Q}}_d(\widetilde{\bm{p}}^{\star}\!
  +\!\Delta\widetilde{\bm{p}})\Vert^2\!+\! \sigma_r^2\text{tr}\big(\bm{W}_{\text{rb}}\bm{W}_{\text{rb}}^{\rm H}\big)
  \le P_R^{\rm max} \! -\! P_J^{\rm max}\vert h_{JR}\vert^2\Vert \bm{W}_{\text{rb}}
  \widetilde{\bm{Q}}_j(\widetilde{\bm{p}}^{\star}\!+\!\Delta\widetilde{\bm{p}})\Vert^2 ,
  \forall\Delta\widetilde{\bm{p}} ,
 \bm{h}_{RE}\bm{R}_{\text{cor}}^{\frac{1}{2}}\bm{W}_{\text{rb}}\widetilde{\bm{Q}}_d\! =\! \bm{0} ,
\end{array}
\end{align}
\hrulefill
\vspace*{-4mm}
\end{figure*}

\begin{figure*}[tp]\setcounter{equation}{91}
\vspace*{-2mm}
\begin{align} 
 &\min\limits_{\Delta\widetilde{\bm{p}}\in{\cal P}} \! \big(\widetilde{\bm{p}}^{\star}\!
  +\!\Delta\widetilde{\bm{p}}\big)^{\rm H}\! \big(\! P_S\vert h_{SR}\vert^2\widehat{\widetilde{\bm{R}}}_d\!-\!
  a P_J^{\rm max}\vert h_{JR}\vert^2\widehat{\widetilde{\bm{R}}}_j\big)\big(\widetilde{\bm{p}}^{\star}\!
  +\!\Delta\widetilde{\bm{p}}\big)\ge a\big(\sigma_r^2\Vert\bm{h}_{RD}^{\rm T}\bm{R}_{\text{cor}}^{\frac{1}{2}}
  \bm{W}_{\text{rb}}\Vert^2\!+\!\sigma_d^2\big) \! , \!  \label{f32} \\ 
 & \max\limits_{\Delta\widetilde{\bm{p}}\in{\cal P}} \big(\widetilde{\bm{p}}^{\star}\!+\!\Delta\widetilde{\bm{p}}\big)^{\rm H}
  \big(P_S\vert h_{SR}\vert^2\widetilde{\bm{Q}}_d^{\rm H}\bm{W}_{\text{rb}}^{\rm H}\bm{W}_{\text{rb}}
  \widetilde{\bm{Q}}_d\!+\! P_J^{\rm max}\vert h_{JR}\vert^2\widetilde{\bm{Q}}_j^{\rm H}
  \bm{W}_{\text{rb}}^{\rm H}\bm{W}_{\text{rb}}\widetilde{\bm{Q}}_j\big)
  \big(\widetilde{\bm{p}}^{\star}\!+\!\Delta\widetilde{\bm{p}}\big) \le P_R^{\rm max}\!-\!
 \sigma_r
 ^2\text{tr}\big(\bm{W}_{\text{rb}}\bm{W}_{\text{rb}}^{\rm H}\big) \label{f33}. 
\end{align}
\hrulefill
\vspace*{-4mm}
\end{figure*}

\subsubsection{Optimization of $\widetilde{\bm{p}}$}

 Given $\bm{W}^{(l)}$, the optimization problem (\ref{subrelay}) is rewritten as\setcounter{equation}{79}
\begin{align}\label{optpsub} 
\begin{array}{l}
 \max\limits_{\widetilde{\bm{p}}} ~  1\! +\! \frac{P_S\vert h_{SR}\vert^2\widetilde{\bm{p}}^{\rm H}
  \widetilde{\bm{Q}}_d^{\rm H}\big(\bm{W}^{(l)}\big)^{\rm H}\big(\bm{R}_{\text{cor}}^{\frac{1}{2}}\big)^{\rm H}
  \bm{h}_{RD}^{\rm H}\bm{h}_{RD}\bm{R}_{\text{cor}}^{\frac{1}{2}} \bm{W}^{(l)}\widetilde{\bm{Q}}_d\widetilde{\bm{p}}}
  {\sigma_d^2+\sigma_r^2\Vert\bm{h}_{RD}\bm{R}_{\text{cor}}^{\frac{1}{2}}\bm{W}^{(l)}\Vert^2} , \\
 \mathrm{s.t.} ~  \text{tr}\big(\widetilde{\bm{p}}^{\rm T}\widetilde{\bm{F}}_n\widetilde{\bm{p}}\big) = 1 ,
  \, 0\le n\le N_R-1 , \\
 ~~~~~ \bm{h}_{RD}\bm{R}_{\text{cor}}^{\frac{1}{2}} \bm{W}^{(l)}\widetilde{\bm{Q}}_j\widetilde{\bm{p}}=0,
  \, \bm{h}_{RE}\bm{R}_{\text{cor}}^{\frac{1}{2}}\bm{W}^{(l)} \widetilde{\bm{Q}}_d\widetilde{\bm{p}}=0 , \\
 ~~~~~ P_S\vert h_{SR}\vert^2\Vert \bm{W}^{(l)}\widetilde{\bm{Q}}_d\widetilde{\bm{p}}\Vert^2 \! + \!
  P_J^{\rm max}\vert h_{JR}\vert^2\Vert \bm{W}^{(l)}\widetilde{\bm{Q}}_j\widetilde{\bm{p}}\Vert^2\\
 ~~~~~~~~~ + \sigma_r^2\text{tr}(\bm{W}^{(l)}\big(\bm{W}^{(l)}\big)^{\rm H}\big)\le P_R^{\rm max} \! . \!
\end{array}
\end{align}
 To simplify this complicated nonconvex problem, we define
\begin{align}\label{eq81}
 \bm{P}_{je}^{(l)} =& \left[\begin{array}{c}
  \bm{h}_{RD}\bm{R}_{\text{cor}}^{\frac{1}{2}}\bm{W}^{(l)}\widetilde{\bm{Q}}_j \\
  \bm{h}_{RE} \bm{R}_{\text{cor}}^{\frac{1}{2}}\bm{W}^{(l)}\widetilde{\bm{Q}}_d
 \end{array} \right]\in \mathbb{C}^{2\times 3N_R} .
\end{align}
 Then the feasible $\widetilde{\bm{p}}$ must be in the null-space of $\bm{P}_{je}^{(l)}$.
 Therefore, $\widetilde{\bm{p}}=\bm{P}_{je}^{{(l)}{\bot}}\widehat{\widetilde{\bm{p}}}$, where
 $\bm{P}_{je}^{{(l)}{\bot}}$ denotes the projection matrix onto the null space of $\bm{P}_{je}^{(l)}$
 and $\widehat{\widetilde{\bm{p}}}$ is the equivalent variable vector to be optimized.
 By defining the Hermitian matrix $\widetilde{\bm{P}}_c=\widehat{\widetilde{\bm{p}}}
 \widehat{\widetilde{\bm{p}}}^{\rm H}$, the problem (\ref{optpsub}) is rewritten as
\begin{align}\label{optpsub2} 
\begin{array}{cl}
 \max\limits_{\widetilde{\bm{P}}_c} & P_S\vert h_{SD}\vert^2
  \text{tr}\big(\widetilde{\bm{R}}_d^{(l)}\widetilde{\bm{P}}_c\big) , \\
 \text{s.t.} & \text{tr}\Big(\widehat{\widetilde{\bm{F}}}_n^{(l)}\! \widetilde{\bm{P}}_c\Big)\! =\! 1 ,
  0\! \le\!  n\! \le\!  N_R\! -\! 1, \\
 & \widetilde{\bm{P}}_c\! \succeq \! \bm{0}, ~ \text{rank}\big(\widetilde{\bm{P}}_c\big )\!  =\! 1 , \\
 & \text{tr}\left(\big(P_S\vert h_{SR}\vert^2\widetilde{\bm{G}}_d^{(l)}\!  +\! P_J^{\rm max}
  \vert h_{JR}\vert^2 \widetilde{\bm{G}}_j^{(l)}\big)\widetilde{\bm{P}}_c\right) \\
 &~~ \le P_R^{\max} - \sigma_r^2\text{tr}\big(\bm{W}^{(l)}{\bm{W}^{(l)}}^{\rm H}\big) ,
\end{array}
\end{align}
 where
\begin{align}
 \widetilde{\bm{R}}_d^{(l)} =& \big(\bm{P}_{je}^{{(l)}{\bot}}\big)^{\rm H}\widetilde{\bm{Q}}_d^{\rm H}
  \big({\bm{W}^{(l)}}\big)^{\rm H}\big(\bm{R}_{\text{cor}}^{\frac{1}{2}}\big)^{\rm H}
  \bm{h}_{RD}^{H}\bm{h}_{RD} \nonumber \\ & \times \bm{R}_{\text{cor}}^{\frac{1}{2}}
  \bm{W}^{(l)}\widetilde{\bm{Q}}_d\bm{P}_{je}^{{(l)}{\bot}}, \label{eq83} \\
 \widehat{\widetilde{\bm{F}}}_n =& \big(\bm{P}_{je}^{{(l)}{\bot}}\big)^{\rm H}\widetilde{\bm{F}}_n
  \bm{P}_{je}^{{(l)}{\bot}} , \label{eq84} \\
 \widetilde{\bm{G}}_k^{(l)} =& \big(\bm{P}_{je}^{{(l)}{\bot}}\big)^{\rm H}\widetilde{\bm{Q}}_k^{\rm H}
  \big(\bm{W}^{(l)}\big)^{\rm H}\bm{W}^{(l)}\widetilde{\bm{Q}}_k\bm{P}_{je}^{{(l)}{\bot}}, \, k=d,j. \label{eq85}
\end{align}
 It is observed that the problem (\ref{optpsub2}) becomes a standard SDP problem if
 the rank-1 constraint is not considered. Similar to solving (\ref{optp}), we utilize the
 penalty based method to solve (\ref{optpsub2}). The corresponding iterative optimization
 problem \eqref{optpsub3} is given at the bottom of this page, where the superscript $^{[t]}$
 denotes the inner iteration index, and $\widetilde{\bm{\upsilon}}_{\max}^{[t]}$ is the
 eigenvector corresponding to the maximum eigenvalue $\lambda_{\max}\big(
 \widetilde{\bm{P}}_c^{[t]}\big)$. Note that the initial $\widetilde{\bm{P}}_c^{[0]}$ and
 the upper bound $\widetilde{\gamma}_{\rm up}
 =\text{tr}\big(\widetilde{\bm{R}}_d^{(l)}\widetilde{\bm{P}}_c^{[0]}\big)$ are derived from
 the problem (\ref{optpsub2}) without considering the rank-1 constraint. Furthermore, the
 penalty based method and the bisection method are jointly applied to iteratively solve
 \eqref{optpsub3} to obtain the optimal rank-1 satisfied $\widetilde{\bm{P}}_c$. This
 procedure is terminated when $\text{tr}\big(\widetilde{\bm{P}}_c^{[t+1]}\big) -
 \lambda_{\max}\big(\widetilde{\bm{P}}_c^{[t+1]}\big)\approx 0$. With the optimal solution
 $\widetilde{\bm{P}}_c^{(l)}=\widetilde{\bm{P}}_c^{[t+1]}$, the optimal $\widetilde{\bm{p}}^{(l)}$
 can be obtained by the eigenvalue decomposition on $\widetilde{\bm{P}}_c^{(l)}$.
 { Since the iterative optimization (\ref{optpsub3}) has exactly the same form as the
 iterative optimization (\ref{optpp}), the convergence of the iterative algorithm
 for solving (\ref{optpsub3}) is guaranteed.}

 Thus, instead of jointly optimizing $\widetilde{\bm{p}}$ and $\bm{W}$, we optimize $\bm{W}$
 and $\widetilde{\bm{p}}$ separately in an iterative procedure involving steps \emph{1)} and
 \emph{2)}. Specifically, with the initial iteration index $l=1$ and a feasible initial
 $\widetilde{\bm{p}}^{(l-1)}$, we obtain the optimal beamforming matrix $\bm{W}^{(l)}$
 using the closed-form solution (\ref{eq77}). Then with $\bm{W}^{(l)}$, the optimal { PSA}
 pointing vector $\widetilde{\bm{p}}^{(l)}$ is determined by solving (\ref{optpsub3})
 iteratively. Because the pair $\big(\bm{W}^{(l)}, \widetilde{\bm{p}}^{(l)}\big)$ is feasible
 in next iteration to obtain $\big(\bm{W}^{(l+1)},\widetilde{\bm{p}}^{(l+1)}\big)$, the
 objective function in the original problem (\ref{subrelay}) is monotonously increasing and
 it converges to the maximum value as the iteration index increases. Hence, when a preset
 termination criterion is met, the procedure yields the optimal beamforming matrix
 $\bm{W}^{\star}$ and the optimal {PSA} pointing $\widetilde{\bm{p}}^{\star}$.

{The complexity of this proposed algorithm mainly comes from the iterative
 SDP optimization (\ref{optpsub3}) for deriving the PSA spatial pointing vector. According to
 \cite{tool}, the computational complexity of a standard SDP problem is on the order of\setcounter{equation}{86}
\begin{align}\label{eq87}
 C_{\rm SDP} \!\!=\! \mathsf{O}\big(\!(M_{sdp}N_{sdp}^{3.5}\!\!+\!\!
  M_{sdp}^2N_{sdp}^{2.5}\!\!+\!\! M_{sdp}^3 N_{sdp}^{0.5})\log(\frac{1}{\epsilon})\! \big) ,
\end{align}
 where $M_{sdp}$ is the number of semidefinite cone constraints and $N_{sdp}$ is the
 dimension of the semidefinite cone, while $\epsilon$ is the accuracy imposed to solve the SDP
 problem. Thus the per-iteration complexity of our proposed algorithm is
\begin{align}\label{eq88}
 C_{\rm per-ite}^{\rm perf} \! \! =\!\! \mathsf{O}\big(\big( (3 N_R)^{3.5}\! \!  +\!\!
  (3 N_R)^{2.5} \!\!  + \! \! (3 N_R)^{0.5} \big) \log(\frac{1}{\epsilon} ) \big).
\end{align}
}\vspace{-5mm}
\subsection{Robust Design for Maximizing Secrecy Rate}\label{S4.2}

 In the previous subsection, we obtain the optimal { PSA} pointing $\widetilde{\bm{p}}^{\star}$.
 In most practical deployments, however, the actual array spatial pointing implemented
 will deviate from this ideal one, due to the antenna distortion, operational
 environment factors or installation errors. Therefore, it is necessary to design a robust
 beamforming under an imperfect { PSA} pointing realization $\widetilde{\bm{p}}_{\text{act}}$.
 There exist two types of array pointing errors. One is modeled as a deterministic matrix
 with bounded norm, and the other is unbounded and denoted by a statistical model of
 unknown parameters. For simplicity,  we only consider the design of robust relay
 beamforming $\bm{W}_{\rm rb}$ for the bounded { PSA} pointing error type. Specifically, an
 ellipsoid model is utilized to model the { PSA} spatial pointing error as \setcounter{equation}{88}
\begin{align}\label{eq89}
 \widetilde{\bm{p}}_{\text{act}} = \widetilde{\bm{p}}^{\star} + \Delta\widetilde{\bm{p}} , \,
 \Delta\widetilde{\bm{p}}\in {\cal P}=\big\{\Delta\widetilde{\bm{p}}: \Delta\widetilde{\bm{p}}^{\rm H}
 \bm{C}\Delta\widetilde{\bm{p}} \le 1 \big\} ,
\end{align}
 where $\widetilde{\bm{p}}^{\star}$ is the optimal { PSA} pointing for the relaying network
 obtained in Subsection~\ref{S4.1}, $\Delta\widetilde{\bm{p}}$ is the elliptical array
 pointing error, and the matrix $\bm{C}\succ 0 $ determines the accuracy degree of the { PSA}
 pointing, which has the Cholesky decomposition of $\bm{C}=\bm{C}^{\frac{1}{2}}
 \big(\bm{C}^{\frac{1}{2}}\big)^{\rm H}$. If the elements of $\bm{C}$ tend to infinity, the
 { PSA} pointing error approaches zero, i.e., the actual array structure is perfect. On the
 other hand, if the elements of $\bm{C}$ approach 0, the { PSA} pointing is extremely inaccurate.

 By considering the array pointing error, the resulting robust secrecy rate maximization
 problem \eqref{f24} is formulated at the top of this page. This optimization is
 highly complicated, and we make some operational assumptions in order to simplify it.
 First, the robust beamforming matrix $\bm{W}_{\text{rb}}$ is designed to satisfy
 $\bm{h}_{RE}^{\rm T}\bm{R}_{\text{cor}}^{\frac{1}{2}}\bm{W}_{\text{rb}}\widetilde{\bm{Q}}_d=\bm{0}$
 to cancel the information leakage from $R$ to $E$ completely. Second, $P_J^{(1)}=P_J^{\rm max}$
 and $P_J^{(2)}=0$ are also applied to the robust beamforming optimization with the same
 reasons as given in Subsection~\ref{S4.1}. Under these conditions, \eqref{f24}
 can be reformulated as the optimization problem \eqref{f25} given at the top of this page,
 where 
 $\widehat{\widetilde{\bm{R}}}_{k} = (\bm{h}_{RD}^{\rm T}\bm{R}_{\text{cor}}^{\frac{1}{2}}\bm{W}_{\text{rb}}
  \widetilde{\bm{Q}}_k)^{\rm H}\bm{h}_{RD}^{\rm T}\bm{R}_{\text{cor}}^{\frac{1}{2}}\bm{W}_{\text{rb}}
 \widetilde{\bm{Q}}_k , \, k=d,j , $
 and the lower bound of $\bar{\gamma}$ is zero, while the upper bound  of $\bar{\gamma}$
 is calculated based on the optimal $\widetilde{\bm{p}}^{\star}$ and $\bm{W}^{\star}$
 from the perfect CSI case. The first two constraints in the optimization problem \eqref{f25}
 can be expressed as \eqref{f32} and \eqref{f33}, respectively, given at the top of this
 page, where 
   $a = 2^{\bar{\gamma}}(1+{P_S\vert h_{SE}\vert^2}/({\sigma_e^2+P_J^{\rm max}\vert h_{JE}\vert^2})) - 1 .$
 To promote the standard SDP formulation for the robust secrecy beamforming design, we
 employ the S-procedure lemma \cite{sp25} to transform  \eqref{f32} and \eqref{f33} into
 the linear matrix inequalities and, consequently, the optimization problem
 (\ref{f25}) is reformulated as \setcounter{equation}{93}
\begin{align}\label{optt25} 
\begin{array}{l}
 \max\limits_{\bm{W}_{\text{rb}},u_1,u_2} ~ \bar{\gamma} , \\
 \mathrm{s.t.} ~ \left[\begin{array}{cc}
  u_1\bm{C}+\bm{\Phi}_{PJ} & \bm{\Phi}_{PJ}^{\rm H}\widetilde{\bm{p}}^{\star} \\
  \big({\widetilde{\bm{p}}^{\star}}\big)^{\rm H}\bm{\Phi}_{PJ} & t_1\end{array} \right] \succeq 0 , \\
 ~~~~~ \left[ \begin{array}{cc}
  u_2\bm{C}-\bm{\Phi}_{JD} & -\bm{\Phi}_{JD}^{\rm H}\widetilde{\bm{p}}^{\star} \\
  -\big(\widetilde{\bm{p}}^{\star}\big)^{\rm H}\bm{\Phi}_{JD} & t_2
  \end{array} \right] \succeq 0 , \\
 ~~~~~ \bm{h}_{RE}\bm{R}_{\text{cor}}^{\frac{1}{2}}\bm{W}_{\text{rb}}\widetilde{\bm{Q}}_d=\bm{0} , ~
  u_1 > 0 , ~ u_2 > 0,
\end{array}
\end{align}
 where
\begin{align}
 & t_1 = \big({\widetilde{\bm{p}}^{\star}}\big)^{\rm H}
  \bm{\Phi}_{PJ}\widetilde{\bm{p}}^{\star} - u_1 - \sigma_r^2 a\big(1\! +\! \Vert\bm{h}_{RD}^{\rm T}
  \bm{R}_{\text{cor}}^{\frac{1}{2}}\bm{W}_{\text{rb}}\Vert^2\big) ,  \nonumber\\
 & t_2 = P_R^{\rm max} - u_2 -
  \big({\widetilde{\bm{p}}^{\star}}\big)^{\rm H}\bm{\Phi}_{JD}\widetilde{\bm{p}}^{\star}
  -\sigma_r^2\text{tr}\big(\bm{W}_{\text{rb}}\bm{W}_{\text{rb}}^{\rm H}\big) , \nonumber \\
 & \bm{\Phi}_{PJ} \!= \!P_S\vert h_{SD}\vert^2\widehat{\widetilde{\bm{R}}}_{d} \! - \! a P_J^{\rm max}
  \vert h_{JR}\vert^2\widehat{\widetilde{\bm{R}}}_{j}, \\
   & \bm{\Phi}_{JD} \!= \!P_S\vert h_{SR}\vert^2\widetilde{\bm{Q}}_d^{\rm H}\bm{W}_{\text{rb}}^{\rm H}
  \bm{W}_{\text{rb}}\widetilde{\bm{Q}}_d \! \!+\! \!P_J^{\rm max}\vert h_{JR}\vert^2\widetilde{\bm{Q}}_j^{\rm H}
  \bm{W}_{\text{rb}}^{\rm H}\bm{W}_{\text{rb}}\widetilde{\bm{Q}}_j , \nonumber
\end{align}
The optimization (\ref{optt25}) is still nonconvex with respect to
 $\bm{W}_{\text{rb}}$. Similar to solving (\ref{optpsub}), some mathematical
 transformations are applied to reformulate (\ref{optt25}) into a standard SDP problem
 with  a rank-1 constraint. Specifically, by defining
\begin{align}\label{eq101}
 \bm{W}_{\text{rb}}^{'} =&\text{vec}(\bm{W}_{\text{rb}})\text{vec}(\bm{W}_{\text{rb}})^{\rm H}
  \in\mathbb{C}^{N_R^2\times N_R^2} ,
\end{align}
 the problem (\ref{optt25}) is transformed into
\begin{align}\label{eq102}
\begin{array}{l}
 \max\limits_{\bm{W}_{\text{rb}}^{'},u_1,u_2} ~ \bar{\gamma}, \\
 \mathrm{s.t.} ~ \left[\begin{array}{cc}
  u_1\bm{C}+\bm{\Phi}_{PJ}^{'} & \bm{\Phi}_{PJ}^{'{\rm H}}\widetilde{\bm{p}}^{\star} \\
  \big({\widetilde{\bm{p}}^{\star}}\big)^{\rm H}\bm{\Phi}_{PJ}^{'} & t_1^{'}
  \end{array}\right] \succeq 0 , \\
 ~~~~~ \left[\begin{array}{cc} u_2\bm{C}-\bm{\Phi}_{JD}^{'} & -\bm{\Phi}_{JD}^{'{\rm H}}\widetilde{\bm{p}}^{\star} \\
  -\big({\widetilde{\bm{p}}^{\star}}\big)^{\rm H}\bm{\Phi}_{JD}^{'} & t_2^{'}
  \end{array} \right] \succeq 0 , \\
 \hspace*{7mm} \text{tr}\big(\widehat{\widetilde{\bm{Q}}}_e\bm{W}_{\text{rb}}^{'}\big)=0, ~ \bm{W}_{\text{rb}}^{'}
 \succeq 0,~\text{rank}(\bm{W}_{\text{rb}}^{'})=1 , \\
 \hspace*{7mm} u_1>0,~ u_2>0,
\end{array}
\end{align}
 where
\begin{align}
 & t_1^{'} = \big({\widetilde{\bm{p}}^{\star}}\big)^{\rm H}
  \bm{\Phi}_{PJ}^{'}\widetilde{\bm{p}}^{\star} - u_1 - \sigma_r^2
  a\left(1+\text{tr}\big(\bm{G}_b\bm{W}_{\text{rb}}^{'}\big)\right) , \nonumber \\
 & t_2^{'} = P_R^{\rm max} - u_2 -
  \big({\widetilde{\bm{p}}^{\star}}\big)^{\rm H}\bm{\Phi}_{JD}^{'}
  \widetilde{\bm{p}}^{\star} - \sigma_r^2\text{tr}\big(\bm{W}_{\text{rb}}^{'}\big) , \nonumber \\
 &\bm{\Phi}_{PJ}^{'} \!= \! P_S\vert h_{SD}\vert^2\widehat{\widetilde{\bm{Q}}}_d^{\rm H}
  \big(\bm{W}_{\text{rb}}^{'}\big)^{\ast}\widehat{\widetilde{\bm{Q}}}_d\!-\!a P_J^{\rm max}
  \vert h_{JR}\vert^2\widehat{\widetilde{\bm{Q}}}_j^{\rm H}\big(\bm{W}_{\text{rob}}^{'}\big)^{\ast}
  \widehat{\widetilde{\bm{Q}}}_j , \nonumber \\
 &\widehat{\widetilde{\bm{Q}}}_k = \big(\widetilde{\bm{Q}}_k\otimes\big(\bm{R}_{\text{cor}}^{\frac{1}{2}}\big)^{\rm T}
  \bm{h}_{RD}\big),~k=d,j,  \\
& \bm{\Phi}_{JD}^{'} = P_S\vert h_{SR}\vert^2\widetilde{\bm{D}}_d+P_J^{\rm max}\vert h_{JR}\vert^2
  \widetilde{\bm{D}}_j ,\nonumber  \\
 &\widehat{\widetilde{\bm{Q}}}_e = \big(\widetilde{\bm{Q}}_d^{\ast}\otimes
  \big(\bm{h}_{RE}^{\rm T}\bm{R}_{\text{cor}}^{\frac{1}{2}}\big)^{\rm H}\big)\big(\widetilde{\bm{Q}}_d^{\rm T}
  \otimes \bm{h}_{RE}^{\rm T}\bm{R}_{\text{cor}}^{\frac{1}{2}}\big) , \nonumber\\
& \widetilde{\bm{D}}_k[\widehat{m},\widehat{n}] = \text{tr}\big(\big(\widetilde{\bm{Q}}_k^{\ast} \otimes
  \bm{I}_{N_R}\big) \widetilde{\bm{I}}_{\widehat{m}}^{\rm H}\widetilde{\bm{I}}_{\widehat{n}}
  \big(\widetilde{\bm{Q}}_k^{\rm T}\!\otimes\!\bm{I}_{N_R}\big)\bm{W}_{\text{rb}}^{'}\big) , k=d,j, \nonumber
\end{align}
 with $\widetilde{\bm{I}}_l = \big[\bm{0}_{N_R\times(l-1)N_R} ~ \bm{I}_{N_R} ~ \bm{0}_{N_R\times
 (3N_R-l)N_R} \big]\in\mathbb{C}^{N_R\times 3N_R^2}$, for $l=\widehat{m},\widehat{n}$ and $1\le
 \widehat{m},\widehat{n}\le 3N_R$. The penalty based method can also be used to solve the problem
 (\ref{eq102}) effectively, and the obtained $\bm{W}_{\text{rb}}^{'\star}$ guarantees to satisfy
 the rank-1 property. The detailed optimization procedure is the same as that presented in
 Subsection~\ref{S4.1}. {This proposed robust beamforming algorithm is based on the iterative
 SDP optimization, which is similar to the one we used to solve the secrecy rate maximization
 with perfect  CSI presented in Subsection~\ref{S4.1}. Therefore, it has the same order of
 magnitude of the per-iteration complexity as the algorithm of Subsection~\ref{S4.1}, which
 can be shown to be $
 C_{\rm per-ite}^{\rm robu} = \mathsf{O}\big(2(N_R^2)^{3.5}+4
  (N_R^2)^{2.5}+ 8(N_R^2)^{0.5})\log(1/\epsilon) \big).$
}

\begin{table*}[tp!]
\begin{small}
\begin{center}
\caption{The list of major simulation variables and optimized results.}
\label{TAB1}
\begin{tabular}{|l|l|l|}
 \hline
  Scenario & DOA of $s_j$ & Optimized 8-element array spatial angles $\big(\bm{\theta}^{(e)},\bm{\varphi}^{(e)}\big)$
  (degrees) for given $\Delta_p$ \\
 \hline
 \multirow{6}{*}{SIMO Network} & $\big(35^\circ,90^\circ\big)$ & \scriptsize{$\!\left[\!\!\!\!\begin{array}{c}
  \big(\bm{\theta}^{(e)}\big)^{\rm T} \\ \big(\bm{\varphi}^{(e)}\big)^{\rm T} \end{array}\!\!\!\!\right]\!
  =\left\{ \!\!\! \begin{array}{ll}
  \Big[\!\!\begin{array}{rrrrrrrr}
   17.24 & 136.24 & 158.07 & 40.95 & 28.56 & 142.68 & 148.61 & 37.59 \\
   73.61 & -51.26 & 73.60 & -52.79 & 83.83 & -52.79 & 87.36 & -52.06 \end{array}\!\!\Big] & \!\!\Delta_p=20^\circ \!\!\!\! \\
  \Big[\!\!\begin{array}{rrrrrrrr}
   109.87 & 107.99 & 66.16 & 19.03 & 142.15 & 143.02 & 42.10 & 52.63 \\
   -60.87 & -44.87 & -67.54 & -17.06 & 56.72 & 64.35 & -19.63 & 58.86 \end{array}\!\!\Big] & \!\!\Delta_p=10^\circ \!\!\!\! \\
  \Big[\!\!\begin{array}{rrrrrrrr}
   87.41 & 88.17 & 90.56 & 90.35 & 88.85 & 90.81 & 90.76 & 90.69 \\
   45.17 & 45.17 & 45.17 & 45.17 & 45.17 & 45.17 & 45.17 & 45.17 \end{array}\!\!\Big] & \!\!\Delta_p=0^\circ \!\!\!\!
  \end{array}\right.$} \\
  \cline{2-3} & $\big(55^\circ,90^\circ\big)$ & \scriptsize{\!$\left[\!\!\!\!\begin{array}{c}
  \big(\bm{\theta}^{(e)}\big)^{\rm T} \\ \big(\bm{\varphi}^{(e)}\big)^{\rm T} \end{array}\!\!\!\!\right]\!
  =\left\{ \!\!\! \begin{array}{ll}
  \Big[\!\!\begin{array}{rrrrrrrr}
   65.67 & 54.43 & 125.57 & 118.73 & 58.17 & 66.55 & 114.63 & 123.94 \\
   -66.74 & -59.71 & -66.31 & -54.00 & -54.45 & -63.38 & -57.70 & -69.35 \end{array}\!\! \Big] & \!\!\Delta_p=20^\circ \!\!\!\! \\
  \Big[\!\!\begin{array}{rrrrrrrr}
   65.71 & 55.94 & 124.14 & 119.30 & 58.59 & 65.80 & 116.15 & 122.38 \\
   -71.70 & -58.84 & -68.75 & -52.39 & -52.65 & -66.24 & -57.48 & -74.12 \end{array}\!\! \Big] & \!\!\Delta_p=10^\circ \!\!\!\! \\
  \Big[\!\!\begin{array}{rrrrrrrr}
   71.47 & 44.56 & 90.00 & 90.00 & 90.00 & 90.00 & 135.44 & 108.53 \\
   63.16 & -55.80 & 85.27 & -89.71 & 20.11 & 25.30 & -55.80 & 63.17 \end{array}\!\!\Big] & \!\!\Delta_p=0^\circ \!\!\!\!
  \end{array}\right.$} \\
  \hline
  \multirow{6}{*}{Relaying Network} & $\Big(35^\circ,90^\circ\big)$ & \scriptsize{$\! \left[\!\!\!\!\begin{array}{c}
   \big(\bm{\theta}^{(e)}\big)^{\rm T} \\ \big(\bm{\varphi}^{(e)}\big)^{\rm T} \end{array}\!\!\!\!\right]\!
   =\left\{ \!\!\! \begin{array}{ll}
  \Big[\!\!\begin{array}{rrrrrrrr}
   92.18 & 91.54 & 91.88 & 96.72 & 90.66 & 81.99 & 93.32 & 91.50 \\
   36.02 & 36.42 & 36.13 & 36.00 & 36.03 & 36.07 & 36.56 & 36.74 \end{array}\!\!\Big] & \!\! \Delta_p=20^\circ \!\!\!\! \\
  \Big[\!\!\begin{array}{rrrrrrrr}
   94.05 & 85.94 & 94.05 & 85.94 & 85.94 & 85.94 & 94.05 & 85.94 \\
   34.02 & 34.03 & 34.07 & 34.02 & 34.03 & 34.02 & 34.02 & 34.02 \end{array}\!\!\Big] & \!\! \Delta_p=10^\circ \!\!\!\! \\
  \Big[\!\!\begin{array}{rrrrrrrr}
   84.10 & 84.97 & 84.73 & 86.87 & 94.99 & 91.28 & 95.47 & 88.87 \\
   30.13 & 30.12 & 30.12 & 30.12 & 30.12 & 30.11 & 30.12 & 30.11 \end{array}\!\!\Big] & \!\! \Delta_p=0^\circ \!\!\!\!
  \end{array}\right.$} \\
  \cline{2-3} &  $\big(55^\circ,90^\circ\big)$ & \scriptsize{\!$\left[\!\!\!\!\begin{array}{c}
  \big(\bm{\theta}^{(e)}\big)^{\rm T} \\ \big(\bm{\varphi}^{(e)}\big)^{\rm T} \end{array}\!\!\!\!\right]\!
  =\left\{ \!\!\! \begin{array}{ll}
  \Big[\!\!\begin{array}{rrrrrrrr}
   114.25 & 129.26 & 62.07 & 63.40 & 118.09 & 113.27 & 64.72 & 70.00 \\
   -42.17 & -43.39 & -44.34 & -50.01 & -50.70 & -44.34 & -46.38 & -41.88 \end{array}\!\!\Big] & \!\! \Delta_p=20^\circ \!\!\!\! \\
  \Big[\!\!\begin{array}{rrrrrrrr}
   64.05 & 65.03 & 116.16 & 116.37 & 64.84 & 64.40 & 114.38 & 114.95 \\
   39.79 & -34.00 & -46.81 & 43.68 & 47.26 & -48.34 & 35.78 & -42.98 \end{array}\!\!\Big] & \!\! \Delta_p=10^\circ \!\!\!\! \\
  \Big[\!\!\begin{array}{rrrrrrrr}
   76.81 & 127.90 & 98.78 & 95.53 & 95.14 & 78.60 & 58.68 & 109.30 \\
   61.70 & -17.16 & -64.91 & 30.57 & -75.27 & -35.15 & 60.26 & -36.25 \end{array}\!\!\Big] & \!\! \Delta_p=0^\circ \!\!\!\!
  \end{array}\right.$} \\
 \hline
\end{tabular}
\end{center}
\end{small}
\vspace{-1mm}
 \end{table*}

\section{Simulation Results}\label{S5}

 In the simulation study, the DOA of desired signal $s_d$ is specified by
 $(\theta_d=40^{\circ},\varphi_d=90^{\circ})$, and its POA is given by
 $(\alpha_d=-30^{\circ}, \beta_d=0^{\circ})$. For jammer signal $s_j$, its DOA is
 $(\theta_j,90^{\circ})$ with $\theta_j\in [0, ~\pi ]$, and its POA is $(\alpha_j,\beta_j)$
 with $\alpha_j \in[-\frac{\pi}{2}, ~ \frac{\pi}{2}]$ and $\beta_j\in[-\frac{\pi}{4},~
 \frac{\pi}{4}]$ . Thus, the spatial and polarization distances between $s_d$ and $s_j$
 are given respectively by $\Delta_a = \vert\theta_j - 40^{\circ} \vert$ and $
 \Delta_p =\arccos \big(\cos2\beta_d\cos2\beta_j\cos\big(2(\alpha_d-\alpha_j)\big)
  +\sin2\beta_d\sin2\beta_j\big)$ \cite{froei}. The 8-antenna {PSA} is considered in
 our simulations and the antenna spacing is half of the transmit signal
 wavelength {\footnote{{Note that our work can also be extended easily to the
 planar array case with the different spatial phase matrix. Moreover, since the  spatial
 phase matrix is not related to any optimization variable, the simulation  conclusions
 obtained by applying the planar array are similar to that  in Section V.}}}. In the
 SIMO network, the eavesdropper is equipped with $N_E=6$ antennas. All  channel coefficients
 are generated independently according to ${\cal{CN}}(0,1)$ and the power of the receive
 additive noise is $\sigma_e^2=\sigma^2=1$. In order to solve the standard convex optimization
 problems such as the GP (\ref{GP2}) and the SDP (\ref{optpsub3}) efficiently, the software
 toolbox CVX \cite{tool} is used. Under this simulation setting, we perform numerical
 evaluations for the point-to-point SIMO network and the relaying network, respectively.
 In order to demonstrate the advantages of {PSA}, the standard CSA based
 technique is utilized as a comparison. { Specifically, instead of employing the 8-element
 PSA, the destination $D$ also employs the CSA with 8 antennas in the SIMO network case,
 while the relay $R$ is also equipped with the 8-element CSA in the relay network case.} All
 the results are averaged over 500 Monte Carlo simulations.

 {For different DOAs of the jammer signal $s_j$, Table~\ref{TAB1} presents
 the optimized spatial pointings of the 8-antenna PSA for both the SIMO network and the
 relaying network is presented. The corresponding optimal destination beamforming and
 relay beamforming can be derived from (\ref{optww}) and (\ref{eq77}), respectively.}

\begin{figure}[t]
\vspace*{-4mm}
\begin{center}
 \includegraphics[width=0.95\columnwidth]{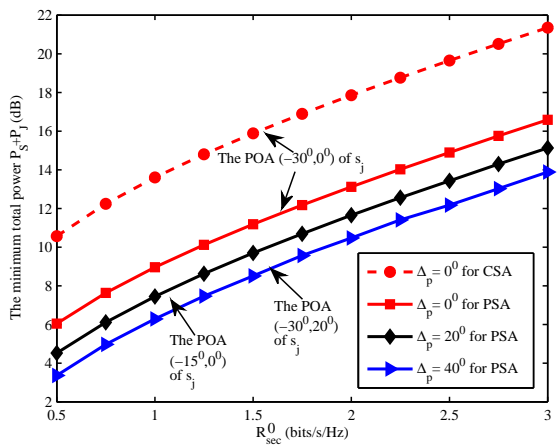}
\end{center}
\vspace*{-6mm}
\caption{The minimum total power consumption as the function of the secrecy rate threshold
 $R_{\rm sec}^0$ under different polarization distances $\Delta_p$. The DOA of $s_j$ is
 $(35^{\circ},90^{\circ})$.}
\label{FIG4}
\vspace*{-5mm}
\end{figure}

\subsection{The SIMO Network}\label{S5.1}

 We first consider the security performance of the SIMO network with the proposed algorithm
 for optimizing the receive beamforming, power allocation and { PSA} spatial pointings.

\subsubsection{Total power minimization}

 With the $8$-antenna { PSA}, Fig.~\ref{FIG4} depicts the minimum power consumption of the
 SIMO network as the function of the secrecy rate threshold $R_{\rm sec}^0$, under three
 different polarization distances $\Delta_p$ with the DOA of jammer signal $s_j$ given by
 $(35^{\circ},90^{\circ})$, which is slightly difficult from the DOA $(40^{\circ},90^{\circ})$
 of desired signal $s_d$. Thus, the spatial distance between $s_d$ and $s_j$ is only
 $\Delta_a=5^{\circ}$, which is considered to be very small. Three POA values considered
 for $s_j$ are $(\alpha_j,\beta_j)\in \{(-30^{\circ},0^{\circ}),(-20^{\circ},0^{\circ})
,(-30^{\circ},20^{\circ})]$, corresponding to three polarization distances $\Delta_p=0^{\circ}$,
 $20^{\circ}$ and $40^{\circ}$, respectively. It is obvious that the minimum total power
 consumption is a monotonously increasing function of the required secrecy rate
 $R_{\text{sec}}^0$, as clearly indicated in Fig.~\ref{FIG4}. Also observe from
 Fig.~\ref{FIG4} that for the { PSA}, increasing $\Delta_p$ leads to reduction in the total
 power consumption, which confirms that the polarization  difference between the two signals
 is beneficial to improve the power efficiency of the { PSA} based SIMO network. As a
 comparison, the minimum total power consumption required by the CSA based SIMO network with
 the POA of $s_j$ given by $(-30^{\circ},0^{\circ})$ is also plotted in Fig.~\ref{FIG4},
 where it can be seen that the CSA needs consume more than 4\,dB of power to achieve the
 required $R_{\text{sec}}^0$, compared with the { PSA}. This is because  the CSA cannot utilize
 the signals' polarization information to improve performance.

\begin{figure}[t]
\vspace*{-7mm}
\begin{center}
\includegraphics[width=0.95\columnwidth]{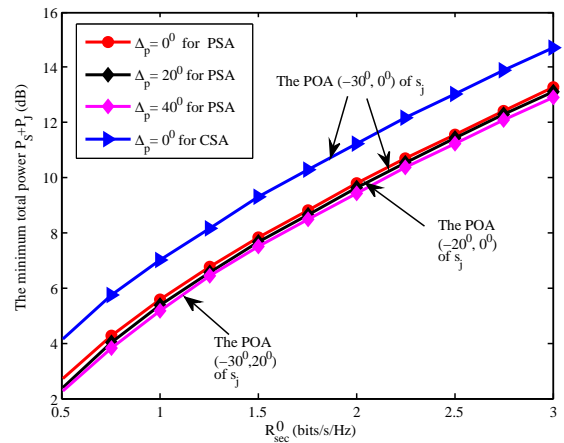}
\end{center}
\vspace*{-6mm}
\caption{The minimum total power consumption as the function of the secrecy rate threshold
 $R_{\rm sec}^0$ under different polarization distances $\Delta_p$. The DOA of $s_j$ is
 $(10^{\circ},90^{\circ})$.}
\label{FIG5}
\vspace*{-1mm}
\end{figure}

\begin{figure}[t]
\vspace*{-7mm}
\begin{center}
\includegraphics[width=0.95\columnwidth]{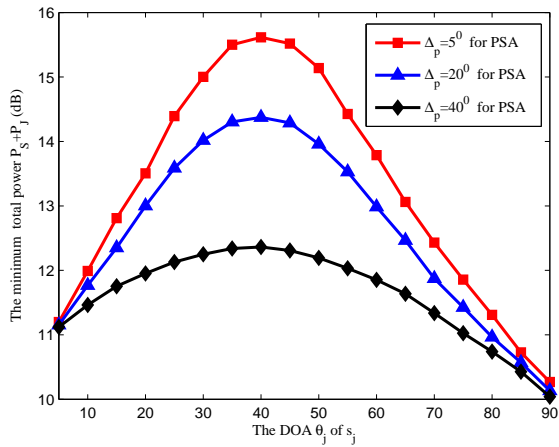}
\end{center}
\vspace*{-6mm}
\caption{The minimum total power consumption as the function of the PSA SIMO network
 as the function of the DOA $(\theta_j,90^{\circ})$ of $s_j$ under different polarization
 distances $\Delta_p$ and given $R_{\rm sec}^0=2.5$\,bits/s/Hz.}
\label{FIG6}
\vspace*{-1mm}
\end{figure}

 We then change the DOA of $s_j$ to $(10^{\circ},90^{\circ})$, and repeat the same
 experiment. The results obtained are shown in Fig.~\ref{FIG5}. Compared with Fig.~\ref{FIG4},
 we observe that the power consumptions of the both CSA-based and { PSA}-based SIMO networks
 are greatly reduced, because we have a large spatial difference $\Delta_a=30^{\circ}$
 between $s_d$ and $s_j$. From Fig.~\ref{FIG5}, it can be seen that the three minimum power
 consumption curves of the { PSA} for the three different polarization distances $\Delta_p$
 become very close. This phenomenon demonstrates that the polarization difference $\Delta_p$
 has little effect on the network performance when the two signals have a sufficiently large
 spatial distance. The results of Fig.~\ref{FIG5} again confirm the advantage of the { PSA} over
 the CSA, as the former achieves 2\,dB saving in power consumption in comparison with the latter.

 Additionally, Fig.~\ref{FIG6} depicts the minimum total power consumption of the PSA SIMO
 network as the function of DOA $(\theta_j,90^{\circ})$ of $s_j$ under three different
 polarization distances $\Delta_p$ and given the secrecy rate threshold $R_{\rm sec}^0=2.5$\,bits/s/Hz.
 It can be seen that as  the DOA difference between $s_d$ and $s_j$, $\Delta_a \to 0$, the
 power consumption reaches the highest value. Again, increasing the polarization distance
 $\Delta_p$ leads to the reduction in power consumption, as clearly shown in Fig.~\ref{FIG6}.
 Furthermore, when the spatial separation $\Delta_a$ is sufficiently large, the influence of
 the polarization difference $\Delta_p$ to power consumption becomes very small.

\begin{figure}[tp!]
\vspace*{-1mm}
\begin{center}
\includegraphics[width=0.95\columnwidth]{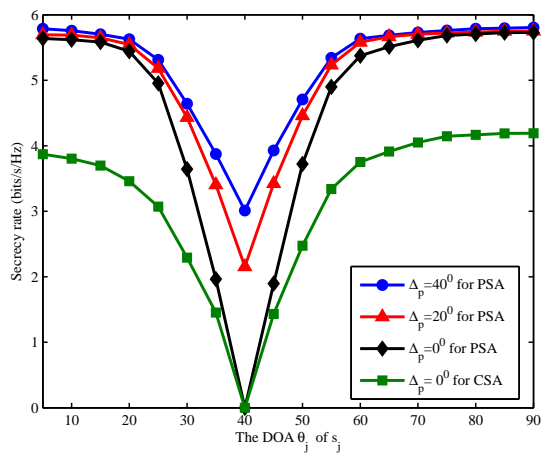}
\end{center}
\vspace*{-6mm}
\caption{The achievable secrecy rate as the function of the DOA $(\theta_j,90^{\circ})$ of
 $s_j$ under different polarization distances $\Delta_p$ and given the total transmit power
 $P_{\text{max}}=12$\,dB.}
\label{FIG7}
\vspace*{-4mm}
\end{figure}

\begin{figure}[tp!]
\vspace*{-5mm}
\begin{center}
\includegraphics[width=0.95\columnwidth]{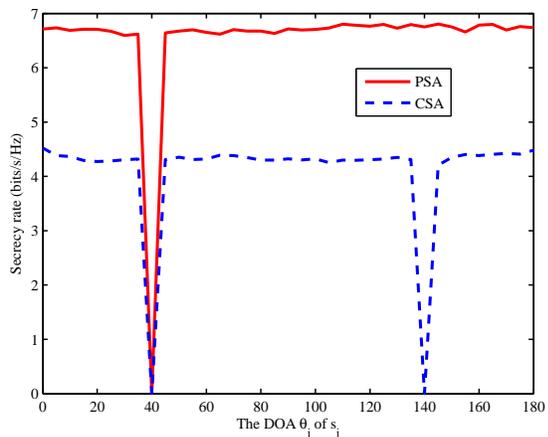}
\end{center}
\vspace*{-6mm}
\caption{The achievable secrecy rate as the function of the DOA $(\theta_j,90^{\circ})$
 of $s_j$ given the polarization distance $\Delta_p=0$ and the total transmit power
 $P_{\text{max}}=14$\,dB. Note that the range of $\theta_j$ is expanded from $90^{\circ}$
 to $180^{\circ}$.}
\label{FIG8}
\vspace*{-1mm}
\end{figure}

\subsubsection{Secrecy rate maximization}

 In this investigation, we set the POA of $s_j$ to $(\alpha_j,0^{\circ})$. The polarization
 distance between $s_d$ and $s_j$ is given by $\Delta_p=2\vert\alpha_j +30^{\circ}\vert$.
 Fig.~\ref{FIG7} depicts the achievable secrecy rates of the { PSA} based SIMO network as the
 functions of the DOA $(\theta_j,90^{\circ})$ of $s_j$, given $P_{\text{max}}=P_S^{\text{max}}
 +P_J^{\text{max}}=12$\,dB. It can be seen that for a given $\Delta_p$, the achievable secrecy
 rate is reduced rapidly as the spatial separation between $s_d$ and $s_j$, $\Delta_a=
 \vert\theta_j-40^0\vert$, decreases, and when $\Delta_a\to 0$, the achievable secrecy rate
 reaches the minimum value. It is also clear that the achievable secrecy rate increases with
 the increase of the polarization separation $\Delta_p$. Moreover, the influence of $\Delta_p$
 to the achievable secrecy rate is particularly strong when the two signals are near spatially
 inseparable, while the influence of $\Delta_p$ becomes very small when the signals are
 sufficiently separable in the spatial domain. Note that at $\Delta_p=0$ and $\Delta_a=0$, the
 secrecy rate is zero. As a comparison, the secrecy rate of the CSA-based SIMO network under
 $\Delta_p=0^{\circ}$ is also given in Fig.~\ref{FIG7}, where it is apparent that the { PSA}
 significantly outperforms the CSA.

 Next we increase $P_{\text{max}}$ to 14\,dB and expand the range of $\theta_j$ from
 $[0^{\circ}, ~90^{\circ}]$ to  $[0^{\circ}, ~180^{\circ}]$. Given $\Delta_p=0$,
 Fig.~\ref{FIG8} compares the achievable secrecy rate of the { PSA} SIMO network with that
 of the CSA SIMO network. As expected, both the {PSA} and CSA attain a zero secrecy rate
 at $\theta_j=40^{\circ}$, as at this point, both $\Delta_a=0$ and $\Delta_p=0$.
 However, it is further noticed that for the CSA, the secrecy rate also deteriorates to
 zero when the DOA of $s_j$ is $(140^{\circ}, 90^{\circ})$. This is owing to the symmetric
 fuzzification and is referred to as grating lobe. By contrast, the secure communication
 of the SIMO network employing { PSA} is realized without introducing grating lobes,
 which is another significant advantage of the {PSA} over the CSA.

\subsection{The Relaying Network}\label{S5.2}

 We now investigate the secure communication of the relay aided network. The DOAs and POAs
 of the incident signals $\widetilde{s}_d$ and $\widetilde{s}_j$ are the same as those given
 in Section~\ref{S5.1} for the SIMO network. We concentrate on the maximum secrecy rate of
 the relaying network obtained by the iterative algorithm proposed in Section~\ref{S4.1},
 assuming a perfect realization of the { PSA} pointing vector, but the robust design with
 imperfect realization of the PSA pointing vector is also studied.
\begin{figure}[t]
\vspace*{-4mm}
\begin{center}
\includegraphics[width=0.95\columnwidth]{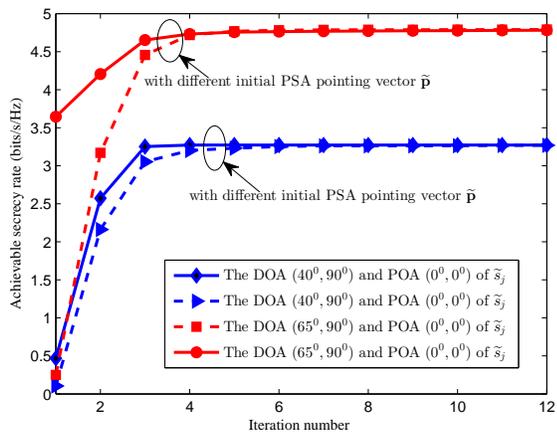}
\end{center}
\vspace*{-6mm}
\caption{Convergence performance of the proposed iterative algorithm for relaying network, given
 $P_s=14$\,dB, $P_R^{\max}=25$\,dB and $P_J^{\max}=10$\,dB.}
\label{FIG9}
\vspace*{-1mm}
\end{figure}

\subsubsection{The secrecy rate maximization for the relaying network}

 First, we demonstrate the convergence of our proposed iterative optimization algorithm
 given in Section~\ref{S4.1}. Specifically, we choose the POA $(0^{\circ}, 0^{\circ})$ for
 $\widetilde{s}_j$, i.e., we consider the case of $\Delta_p=2\vert\alpha_j+30^{\circ}\vert
 =60^{\circ}$, and we set  $P_s=14$\,dB, $P_R^{\max}=25$\,dB and $P_J^{\max}=10$\,dB.
 Fig.~\ref{FIG9} depicts the convergence performance of the proposed iterative optimization
 algorithm under both the spatially separable and spatially inseparable cases. It can be seen
 from Fig.~\ref{FIG9} that for the case of $\Delta_a > 0$, the algorithm takes $l=3$ outer
 iterations to converge, while for the case of $\Delta_a=0$, the algorithm converges within
 $l=6$ outer iterations. Moreover, the choice of the initial $\widetilde{\bm{p}}^{(0)}$
 does not seem to affect the algorithm's convergence performance.

\begin{figure}[tp!]
\vspace*{-1mm}
\begin{center}
\includegraphics[width=0.95\columnwidth]{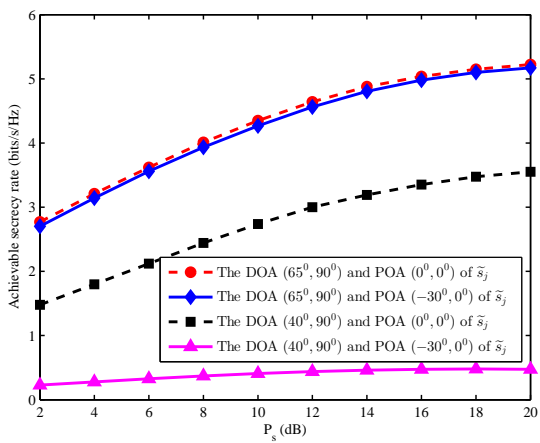}
\end{center}
\vspace*{-6mm}
\caption{The achievable secrecy rate of the { PSA} based relaying network as the function of
 source transmit power $P_S$ under different DOA and POA settings of $\widetilde{s}_j$ with
 $P_J^{\text{max}}=10$\,dB and $P_R^{\text{max}}=25$\,dB.}
\label{FIG10}
\vspace*{-5mm}
\end{figure}

 In Fig.~\ref{FIG10}, the achievable secrecy rate of the { PSA} relaying network is depicted
 as the function of the source transmit power $P_S$, under different DOA $(\theta_j,90^{\circ})$
 and POA $(\alpha_j,0^{\circ})$ of jammer signal $\widetilde{s}_j$ with the maximum relay
 power $P_R^{\rm max}=25$\,dB and the maximum jammer power $P_J^{\rm max}=10$\,dB. Obviously,
 the achievable secrecy rate is a monotonically increasing function of $P_S$ but it exhibits
 a saturation trend for large $P_S$. This is because increasing $P_S$ also increases the
 information leakage from source $S$ to eavesdropper $E$, while the relay transmit power
 $P_R^{\max}$ is limited. Therefore, the achievable secrecy rate cannot go arbitrarily high.
 With the DOA of $\widetilde{s}_j$ given by $(65^{\circ},90^{\circ})$, which is distinguishable
 from the DOA $(40^{\circ},90^{\circ})$ of $\widetilde{s}_d$, the influence of the polarization
 distance $\Delta_p$ between $\widetilde{s}_d$ and $\widetilde{s}_j$ on the achievable secrecy
 rate is very small. However, when $\widetilde{s}_j$ and $\widetilde{s}_d$ are spatially
 inseparable with  $\Delta_a=0^{\circ}$, the influence of $\Delta_p$ becomes significant, and
 a larger $\Delta_p$ leads to a larger achievable secrecy rate. Also observe from Fig.~\ref{FIG10}
 that under the condition of $\Delta_a=0^{\circ}$ and $\Delta_p=0^{\circ}$, the achievable
 secrecy rate of the { PSA} relaying network is very small.

\begin{figure}[t]
\vspace*{-4mm}
\begin{center}
\includegraphics[width=0.95\columnwidth]{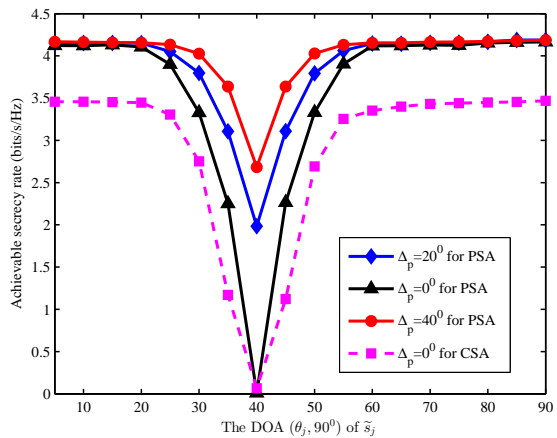}
\end{center}
\vspace*{-6mm}
\caption{The achievable secrecy rate of the relay network as the function of the DOA
 $(\theta_j,90^{\circ})$ of $\widetilde{s}_j$ under different polarization differences
 $\Delta_p$ and given $P_J^{\text{max}}=P_S=10$\,dB and $P_R^{\text{max}}=25$\,dB.}
\label{FIG11}
\vspace*{-2mm}
\end{figure}

 Fig.~\ref{FIG11} compares the achievable secrecy rates of the PSA and CSA relaying networks
 given the DOA $(\theta_j,90^{\circ})$ for $\widetilde{s}_j$ with $\theta_j\in [0^{\circ}, ~
 90^{\circ}]$, $P_J^{\text{max}}=10$\,dB and $P_S=15$\,dB. In this case, the POA of $\widetilde{s}_j$
 is given by $(\alpha_j,0^{\circ})$, and the polarization distance between $\widetilde{s}_d$ and
 $\widetilde{s}_j$ is $\Delta_p=2\vert\alpha_j+30^{\circ}\vert$. The results of  Fig.~\ref{FIG11}
 demonstrate that the {PSA} based relaying network significantly outperforms the CSA  based
 relaying network, in terms of achievable secrecy rate. Similar to the SIMO network case,
 at $\Delta_a=0^{\circ}$, the achievable secrecy rates of both the { PSA} and CSA relaying networks
 deteriorate to their minimum values. In particular, under the condition of $\Delta_a=0^{\circ}$
 and $\Delta_p=0^{\circ}$, the secrecy rate of the CSA relaying network is zero
 but the secrecy rate of the {PSA} relaying network is a small nonzero value. Furthermore, by
 increasing the polarization separation $\Delta_p$ to nonzero, the secrecy rate of the { PSA} relaying
 network can be increased considerably, because the {PSA} can effectively utilize the polarization
 information.

\begin{figure}[t]
\vspace*{-4mm}
\begin{center}
\includegraphics[width=0.95\columnwidth]{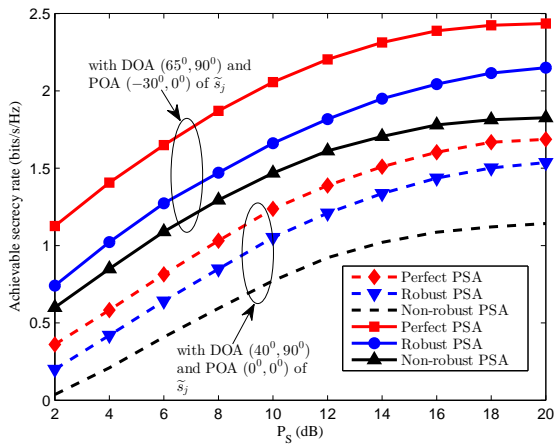}
\end{center}
\vspace*{-6mm}
\caption{The achievable secrecy rates of three designs as the functions of $P_s$,
 under different DOA and POA settings of $\widetilde{s}_j$ with the 4-antenna { PSA} as well
 as $P_J^{\max}=10$\,dB and $P_R^{\max}=25$\,dB.}
\label{FIG12}
\vspace*{-2mm}
\end{figure}

\subsubsection{Robust design with imperfect realization of the { PSA} pointing}

 We next illustrate our robust beamforming optimization design for the { PSA} relaying network
 with imperfect { PSA} pointing realization. We set the { PSA} pointing error bound to
 $\bm{C}=100\bm{I}_{3N_R}$. In order to reduce the computation complexity, we consider the
 $4$-antenna { PSA}. In Fig.~\ref{FIG12}, the achievable secrecy rates of three designs
 as the functions of the maximum source transmit power $P_S$ are depicted, given
 different DOA and POA conditions for $\widetilde{s}_j$ with the 4-antenna { PSA} as well as
 $P_J^{\max}=10$\,dB and $P_R^{\max}=25$\,dB. Based on the secrecy rate maximization of
 Section~\ref{S4.1}, we can obtain the optimal design of $\bm{W}^{\star}$ and
 $\widetilde{\bm{p}}^{\star}$. If the { PSA} pointing implementation is perfect, we can
 realize the exact optimal { PSA} pointing solution $\widetilde{\bm{p}}^{\star}$, which is the
 curve under the title `Perfect {PSA}' in Fig.~\ref{FIG12}. However, in practice, there
 usually exists {PSA} pointing implementation error, and the optimal design of
 $\widetilde{\bm{p}}^{\star}$ is actually implemented as $\widetilde{\bm{p}}^{\star}+
 \Delta\widetilde{\bm{p}}$, which is the curve under the title `Non-robust { PSA}' in
 Fig.~\ref{FIG12}. Obviously, this implementation is far from optimal, and the actual
 secrecy rate achieved is significantly lower than that obtained with the  perfect
 implementation of $\widetilde{\bm{p}}^{\star}$. Under the imperfect implementation of
 $\widetilde{\bm{p}}^{\star}+\Delta\widetilde{\bm{p}}$, our robust beamforming optimization
 design presented in Section~\ref{S4.2} is capable of re-gaining considerable secrecy rate
 performance, which is shown in Fig.~\ref{FIG12} under the title `Robust {PSA}'.
 { \subsection{Extension to the  PSA based eavesdropper }
 As mentioned before, our work can also be easily extended to the case where the PSA is  deployed at the eavesdropper. Note that the  PSA-based eavesdropper can effectively suppress
  the interference with approximate  spatial properties as the source,  and further improve the wiretap capability. In order to illustrate the  good performance of the  PSA-based eavesdropper, we perform the following two simulations for total power minimization  and secrecy rate maximization, respectively.

  Particularly, we assume that the eavesdropper is equipped with a PSA with $N_E=6$ antennas. Besides,
the DOA and POA of the desired signal $s_{d}$ impinging on PSA at eavesdropper are  assumed to be
 $(30^{\circ},90^{\circ})$ and
 $(-30^{\circ}, 0^{\circ})$, respectively. For jammer signal $s_j$, its DOA at eavesdropper  is
 $(25^{\circ},90^{\circ})$ and its POA is $(\alpha_{je},\beta_{je})$. Based on this, the  polarization distance at eavesdropper between $s_d$ and $s_j$
  is defined as  $\Delta_{pe}$, which is similar to that of the polarization distance at destination $\Delta_p$. Note that in the following two
 simulations,   $\Delta_{p}=20^0$ is fixed.
  In Fig.~\ref{FIG13}, the   total power consumption  of the SIMO
network versus  the secrecy rate threshold  $R_{\rm sec}^0$ under  different polarization distances $\Delta_{pe}$ is studied. Here,  both the eavesdropper  equipped with the CSA and PSA (also named CSA-Eve and PSA-Eve), are considered.
Firstly,  we find that for both CSA-Eve and PSA-Eve, the total power  consumption increases with $R_{\rm sec}^0$.
Secondly,  it is clear that
the total power  consumption  for PSA-Eve  case increases with   $\Delta_{pe}$, which dues to the  fact that the wiretap capability of PSA-Eve is enhanced   by enlarging the polarization distance $\Delta_{pe}$.
More importantly,  compared to the case  of CSA-Eve,  the  total power consumption for PSA-Eve case is evidently higher under the arbitrary  polarization
distance $\Delta_{pe}$, which indicates that  more transmit power is required to suppress the interception of PSA-Eve  and further  achieve the secrecy rate threshold.
\begin{figure}[t]
\centering
\includegraphics[width=0.95\columnwidth]{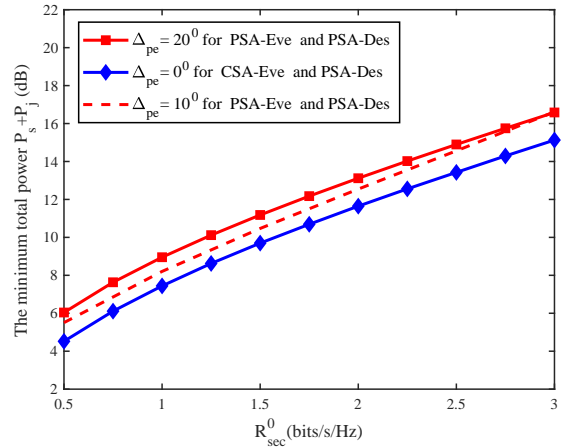}
\caption{The minimum total power consumption as the function of the secrecy rate threshold
 $R_{\rm sec}^0$ under different polarization distances $\Delta_p$. The DOA of $s_{je}$ is
 $(35^{\circ},90^{\circ})$.}
 \label{FIG13}
\end{figure}
  \begin{figure}[t]
\centering
\includegraphics[width=0.95\columnwidth]{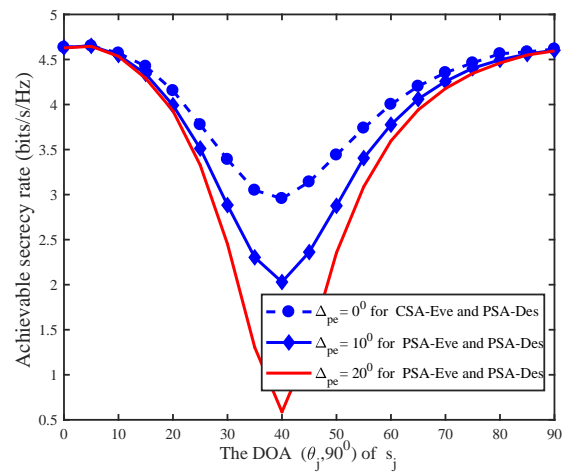}
\caption{The achievable secrecy rate as the function of the DOA $(\theta_{je},90^{\circ})$ of
 $s_{je}$ under different polarization distances $\Delta_{pe}$.} \label{FIG14}
\end{figure}

In Fig.~\ref{FIG14}, the achievable secrecy rate  versus the DOA $(\theta_{j},90^{\circ})$ of  $s_{j}$  for both PSA-Eve and CSA-Eve case is  shown. The same parameters in  Fig. 1  are adopted, and the total   transmit power  $P_S+P_J=12 dB$ is assumed.
Based on this, we naturally find that for both PSA-Eve and CSA-Eve, when the DOA  $(\theta_{j},90^{\circ})$ of $s_{j}$  is near to  that of the desired signal, i.e. $\vert\theta_{j}-40^0\vert<10^o$, the secrecy rate performance evidently deteriorates due to the similar spatial characteristics between the desired signal and the jammer signal.
Besides, with the increase of polarization distance  $\Delta_{pe}$, i.e., from $\Delta_{pe}=10^o$  to $\Delta_{pe}=20^o$, the PSA-Eve can utilize more  polarization difference  to   improve the wiretap rate and  further reduce the achievable secrecy rate.
  More importantly, compared to the case of  CSA-Eve,  the lower achievable secrecy rate of PSA-Eve is observed, which is attributed to    the better wiretap capability of PSA-Eve.}

\section{Conclusions}\label{S6}

 In this paper, a joint beamforming, power allocation and { PSA} pointing optimization
 has been proposed for wireless communications. Our main contribution has been to apply the
 polarization sensitive array to improve the security performance of wireless communications.
 Specifically, by utilizing the polarization difference among signals, the interference
 caused by jammer to destination is greatly reduced while the information leakage to
 eavesdropper is minimized, even when these signals are spatially indistinguishable.  Two
 communication scenarios, the { PSA} based SIMO network and the { PSA} aided relaying network, have
 been considered. For the former scenario, both total transmit power minimization and secrecy
 rate maximization have been performed. For the relaying network assuming perfect CSI, both
 secrecy rate maximization designs under perfect and imperfect { PSA} spatial pointing
 implementations have been obtained. Since all the optimization problems involved are
 nonconvex with complicated constraints and/or objectives, alternative suboptimal algorithms
 have been proposed which enable us to apply convex optimization techniques to solve the
 transformed optimization problems efficiently. Extensive simulation results have demonstrated
 the effectiveness of our proposed { PSA} based techniques for enhancing physical-layer
 security. In particular, it has been shown that the improvement of maximum achievable secrecy
 rate of wireless networks by the proposed { PSA} techniques over the standard CSA techniques
 is remarkable.


{
\appendix

 The convergence of the proposed iterative algorithm to solve the optimization problem
 \eqref{optpp} was demonstrated in \cite{12}. Here we prove the convergence of this
 algorithm.

 Firstly, we define the following objective function
\begin{align}\label{ab1} 
 f({\bm{P}}_c) &=  \text{tr}\big({\bm{P}}_c\big) - \lambda_{\max}\big({\bm{P}}_c\big)
  - \text{tr}\left(\bm{\vartheta}_{\max} \bm{\vartheta}_{\max}^{\rm H}
  \big({\bm{P}}_c - {\bm{P}}_c\big)\right)\nonumber\\
 &= \text{tr}\big({\bm{P}}_c\big) - \lambda_{\max}\big({\bm{P}}_c\big) .
\end{align}
 Then the lower bound of $f({\bm{P}}_c)$ is zero according to the following lemma.
 \vspace{-2mm}
\begin{lemma}\label{L1}
 For an arbitrary square matrix $\bm{A}$, it holds that $\text{tr}(\bm{A})-\lambda_{\max}(\bm{A})
 \ge 0 $, in which the equality is guaranteed if and only if $\text{rank}(\bm{A})=1$ is satisfied.
\end{lemma}
 Next, we  modify $ f({\bm{P}}_c)$ into the following penalty function
\begin{align}\label{ab2} 
 \widetilde{f}(\bm{P}_c^{(t+1)}) =& \min\limits_{\bm{P}_c} ~ \text{tr}({\bm{P}}_c) -
  \lambda_{\text{max}}\big(\bm{P}_c^{(t)}\big) \nonumber \\ & -
  \text{tr}\left(\bm{\vartheta}_{\text{max}}^{(t)}\big(\bm{\vartheta}_{\text{max}}^{(t)}\big)^{\rm H}
  \big({\bm{P}}_c -{\bm{P}}_c^{(t)}\big)\right) ,
\end{align}
 where $\text{tr}(\bm{P}_c) - \lambda_{\text{max}}\big(\bm{P}_c^{(t)}\big) -
 \text{tr}\left(\bm{\vartheta}_{\text{max}}^{(t)}\big(\bm{\vartheta}_{\text{max}}^{(t)}\big)^{\rm H}
  \big({\bm{P}}_c -{\bm{P}}_c^{(t)}\big)\right)$ is the objective function in the
 $(t+1)$th iteration of \eqref{optpp}.

 Let $\bm{P}_c^{(t)}$ be the optimal matrix obtained at the $t$th iteration of \eqref{optpp}.
 Using $\bm{P}_c^{(t)}$ in the iterative optimization procedure yields the optimal
 matrix $\bm{P}_c^{(t+1)}$ at the $(t+1)$th iteration, which is feasible. Clearly,
 the optimized $\widetilde{f}\big(\bm{P}_c^{(t+1)}\big)$ satisfies
\begin{align}\label{ab3} 
 \widetilde{f}\big(\bm{P}_c^{(t+1)}\big) =& \text{tr}\big(\bm{P}_c^{(t+1)}\big) -
  \lambda_{\text{max}}\big(\bm{P}_c^{(t)}\big) \nonumber \\ & -
  \text{tr}\left(\bm{\vartheta}_{\text{max}}^{(t)}\big(\bm{\vartheta}_{\text{max}}^{(t)}\big)^{\rm H}
  \big(\bm{P}_c^{(t+1)}-\bm{P}_c^{(t)}\big)\right) \nonumber \\
 \le & \text{tr}\big({\bm{P}}_c^{(t)}\big) - \lambda_{\text{max}}\big(\bm{P}_c^{(t)}\big) \nonumber \\
 & - \text{tr}\left(\bm{\vartheta}_{\text{max}}^{(t)}\big(\bm{\vartheta}_{\text{max}}^{(t)}\big)^{\rm H}
  \big(\bm{P}_c^{(t)}-\bm{P}_c^{(t)}\big)\right) \nonumber \\
 =& f\big(\bm{P}_c^{(t)}\big) .
\end{align}
 For an arbitrary Hermitian matrix $\bm{Z}$, the following relationship holds
\begin{align}\label{ab4} 
 &\text{tr}\left(\bm{\vartheta}_{\text{max}}\bm{\vartheta}_{\text{max}}^{\rm H}\big(\bm{Z}-\bm{P}_c^{(t)}\big)\right)
 = \bm{\vartheta}_{\text{max}}^{\rm H}\bm{Z}\bm{\vartheta}_{\text{max}}-
  \bm{\vartheta}_{\text{max}}^{\rm H}{\bm{P}}_c^{(t)}\bm{\vartheta}_{\text{max}} \nonumber \\
&\!\! = \!\bm{\vartheta}_{\text{max}}^{\rm H}\bm{Z}\bm{\vartheta}_{\text{max}}\! \!-\!\!
  \lambda_{\max}\big(\bm{P}_c^{(t)}\big)\!\!
 \le\! \! \lambda_{\max}(\bm{Z})\!-\!\lambda_{\max}(\bm{P}_c^{(t)}) .
\end{align}
 Therefore, we further obtain
\begin{align}\label{ab5} 
 & \lambda_{\max}\big({\bm{P}}_c^{(t+1)}\big) - \lambda_{\max}({\bm{P}}_c^{(t)}) \ge
 \nonumber \\ & \hspace*{20mm}
  \text{tr}\left(\bm{\vartheta}_{\text{max}}^{(t)}\big(\bm{\vartheta}_{\text{max}}^{(t)}\big)^{\rm H}
  \big({\bm{P}}_c^{(t+1)}\!\!-\!{\bm{P}}_c^{(t)}\big)\right) \!.
\end{align}
 Based on  \eqref{ab5}, the function $f\big(\bm{P}_c^{(t+1)}\big)$ satisfies
\begin{align}\label{ab6} 
 &f\big(\bm{P}_c^{(t+1)}\big) = \text{tr}\big(\bm{P}_c^{(t+1)}\big) -
  \lambda_{\text{max}}\big(\bm{P}_c^{(t+1)}\big) \nonumber \\
 &\!=\! \text{tr}\big(\bm{P}_c^{(t+1)}\big) \!- \! \lambda_{\text{max}}\big(\bm{P}_c^{(t)}\big)
  \!+\! \lambda_{\text{max}}\big(\bm{P}_c ^{(t)}\big)\!- \! \lambda_{\text{max}}\big(\bm{P}_c^{(t+1)}\big) \nonumber \\
 &\!\le \! \text{tr}\big(\bm{P}_c^{(t+1)}\big)\! - \! \lambda_{\text{max}}\big(\bm{P}_c ^{(t)}\big)\!-\! \text{tr}\left(\bm{\vartheta}_{\text{max}}^{(t)}\big(\bm{\vartheta}_{\text{max}}^{(t)}\big)^{\rm H}
  \big(\bm{P}_c^{(t+1)}\!-\!\bm{P}_c^{(t)}\big)\right) \nonumber \\
 &\le  f\big(\bm{P}_c^{(t)}\big) ,
\end{align}
 where the first inequality  and the second inequality  hold due to \eqref{ab5}
 and \eqref{ab3}, respectively.

 Thus, given an initial feasible ${\bm{P}}_c^{(0)}$, the optimization problem
 \eqref{optpp} can be iteratively solved to obtain a sequence $\bm{P}_c^{(t)}$,
 $t=1,2,\cdots$, whose rank approaches 1. Since this iteration procedure is
 monotonically decreasing, in terms of the objective function $f\big(\bm{P}_c^{(t)}\big)$,
 as shown in (\ref{ab6}), which has a lower bound of zero based on Lemma~\ref{L1},
 it is naturally converged. Consequently, we conclude that the iterative optimization
 problem \eqref{optpp} based on the penalty function method converges to a rank-1
 solution.
}

\end{document}